\documentclass{article}
\usepackage{amsmath}
\usepackage{graphicx}
\usepackage{subcaption}
\usepackage{tabularx}
\usepackage{multirow}
\usepackage{placeins}
\usepackage{enumitem}

\usepackage[numbers, compress]{natbib}
\usepackage[main, final]{neurips_2026}

\usepackage[utf8]{inputenc} 
\usepackage[T1]{fontenc}    
\usepackage{hyperref}       
\hypersetup{
  pdftitle={Breaking Noise Shortcuts in Self-Supervised Learning via Noise-Aligned View Generation},
  pdfauthor={Oankar R. Patil and May D. Wang}
}
\usepackage{url}            
\usepackage{booktabs}       
\usepackage{amsfonts}       
\usepackage{nicefrac}       
\usepackage{microtype}      
\usepackage[dvipsnames]{xcolor}         

\newcommand{\llbracket}{\mathopen{[\mkern-3mu[}}
\newcommand{\rrbracket}{\mathclose{]\mkern-3mu]}}

\newcommand{\pms}[1]{{\scriptsize$\pm$#1}}
\newcommand{\up}[1]{\textcolor{ForestGreen}{+#1}}

\title{Breaking Noise Shortcuts in Self-Supervised Learning via Noise-Aligned View Generation}

\author{%
  Oankar R.~Patil \\
  Department of Electrical \& Computer Engineering\\
  Georgia Institute of Technology\\
  Atlanta, GA, 30332 \\
  \texttt{opatil31@gatech.edu} \\
  \And
  May D. Wang \\
  Department of Biomedical Engineering \\
  Georgia Institute of Technology\\
  Atlanta, GA, 30332 \\
}

\begin{document}

\maketitle

\begin{abstract}
Self-supervised learning (SSL) is increasingly applied to scientific imaging domains, such as microscopy, where labels are scarce but noisy data is abundant. These domains exhibit significant signal-dependent noise, which can create a representational shortcut: since both augmented views derive from the same noisy image, the encoder can boost view agreement by also encoding noise patterns rather than semantic content. To enable noise-robust representation learning without knowledge of underlying noise models, we introduce Self-Aligned Noise Augmentation (SEANA), a drop-in module that generates noise-aware SSL views. SEANA learns an invertible variance-stabilizing transform (VST) from noisy data, then estimates the clean signal and resamples independent noise in learned VST space, requiring no clean targets, no dataset-specific denoiser training or use at inference, and no changes to the SSL objective or encoder. On CIFAR-10 and ImageNet-100 under synthetic Gaussian, Poisson–Gaussian, and multiplicative noise, SEANA improves clean-test linear-probe accuracy across contrastive/non-contrastive SSL methods, indicating higher-quality representations. On real fluorescence imaging, SEANA improves Jurkat cell-cycle stage classification by up to 28.6pp over standard SSL and 27.5pp over denoiser-preprocessed baselines. In all settings, SEANA outperforms the denoiser-preprocessed SSL pipelines, demonstrating that learned VST-space noise-aligned resampling yields more robust representations than denoising alone. Code is available at: \url{https://github.com/opatil31/SEANA}.
\end{abstract}

\section{Introduction}

Self-supervised learning (SSL) learns transferable visual representations without human annotation. By training encoders to produce similar embeddings for augmented image views, SSL methods match or exceed supervised pretraining on many downstream tasks \cite{gui2024survey}. A key assumption underlying this success is that the augmentation pipeline induces variability that is \emph{semantically preserving}, where it perturbs nuisance factors (e.g., cropping, color jitter, blur) while keeping the underlying semantic content intact. In many practical scenarios in scientific imaging, however, there can be severe, signal-dependent \emph{acquisition noise} \cite{foi2009clipped}.

Such noise is ubiquitous. Microscopy and flow-cytometry imagery often exhibits shot noise \cite{petty2007fluorescence}. Coherent imaging introduces speckle noise \cite{bianco2018strategies} and sensor pipelines commonly add approximately Gaussian components. As a result, since the augmented SSL views derive from the same noisy image, the encoder can boost view agreement by also encoding noise patterns rather than semantic content.

Recent theory sharpens this intuition. \citet{van2025joint} analyze SSL under additive nuisance corruption and show that the ability to recover the \emph{noiseless} SSL solution depends on \emph{noise-alignment}. Augmentations must inject (or otherwise induce) variability that is aligned with the nuisance variability present in the data. When the augmentation-induced noise is misaligned, increasing dataset size alone does not guarantee recovery of the clean solution. Instead, one must cross a minimum alignment threshold. While the analysis is developed in a linear-Gaussian setting, it captures a core practical message. If noise is the dominant nuisance factor, SSL benefits from view generation mechanisms that explicitly \emph{resample aligned nuisance} across views. But how can we practically achieve this? A direct application of the noise-alignment prescription would require sampling independent noise from the same conditional distribution of the data corruption. This requires knowledge of the noise model and its parameters, both of which vary across instruments and acquisition pipelines.

As such, in this work, we study joint-embedding SSL in the presence of \emph{signal-dependent, pixelwise independent} noise. Specifically, we investigate synthetic additive Gaussian, Poisson--Gaussian (shot/read), Multiplicative (speckle) noise, and real-world acquisition noise as the primary nuisance.

Thus, to enable noise-robust representation learning without knowledge of the underlying noise model or modifying SSL training curricula, we propose \textbf{Self-Aligned Noise Augmentation (SEANA)}, a drop-in view-generation module for joint-embedding SSL. SEANA learns an invertible, spline-based variance-stabilizing transform (VST) directly from noisy training data, requiring no clean targets. During pretraining, SEANA maps each image to the learned VST space, then uses a frozen, off-the-shelf blind-spot gaussian denoiser to estimate the clean signal, resamples i.i.d.\ Gaussian noise in the VST space, and inverts back to the original intensity space. The resulting views contain \emph{independent} noise realizations that are \emph{aligned} with the original corruption, without requiring knowledge of the underlying noise model. Critically, SEANA leaves the SSL objective and encoder architecture unchanged and requires no dataset-specific denoiser training or denoiser use at downstream inference.
\vspace{-12pt}
\paragraph{Our contributions are as follows:}
\begin{enumerate}[label=\textbf{\arabic*.}, leftmargin=*, noitemsep, topsep=0pt]
    \item We find that for SSL, a learned variance-stabilizing transform (VST) reduces signal-dependent noise to the homoscedastic case where noise-aligned augmentation can be met via simple Gaussian resampling, without knowing the data's underlying noise model or its parameters.
    \item We propose SEANA, a drop-in view-generation module, for noise-robust representation learning in joint-embedding SSL. It learns an invertible VST from noisy data and performs VST-space noise resampling, requiring no clean targets, no dataset-specific denoiser, and no modifications to the SSL objective, encoder, training curricula, or use at inference. 
    \item We demonstrate consistent improvements across SSL methods (BYOL, SimCLR, VICReg, LeJEPA, MoCoV2), three noise families (Gaussian, Poisson--Gaussian, Multiplicative), two synthetic benchmarks (CIFAR-10, ImageNet-100), and two real-world datasets from fluorescence imaging (BBBC048) and clinical ultrasound imaging, with SEANA achieving up to +28.6~pp over baseline SSL and +27.3~pp over denoiser-based SSL in a BBBC048 classification task.
    \item We do an ablation revealing that both the learned VST and noise resampling are essential: resampling without the VST collapses BYOL and hurts MoCoV2, while the VST without resampling matches baseline. SEANA is the only configuration that works universally across SSL methods, while being ${\sim}10{,}000 \times$ more parameter-efficient than training a dataset-specific denoiser.
\end{enumerate}

\begin{figure}[t]
\centering
\includegraphics[width=0.9\textwidth]{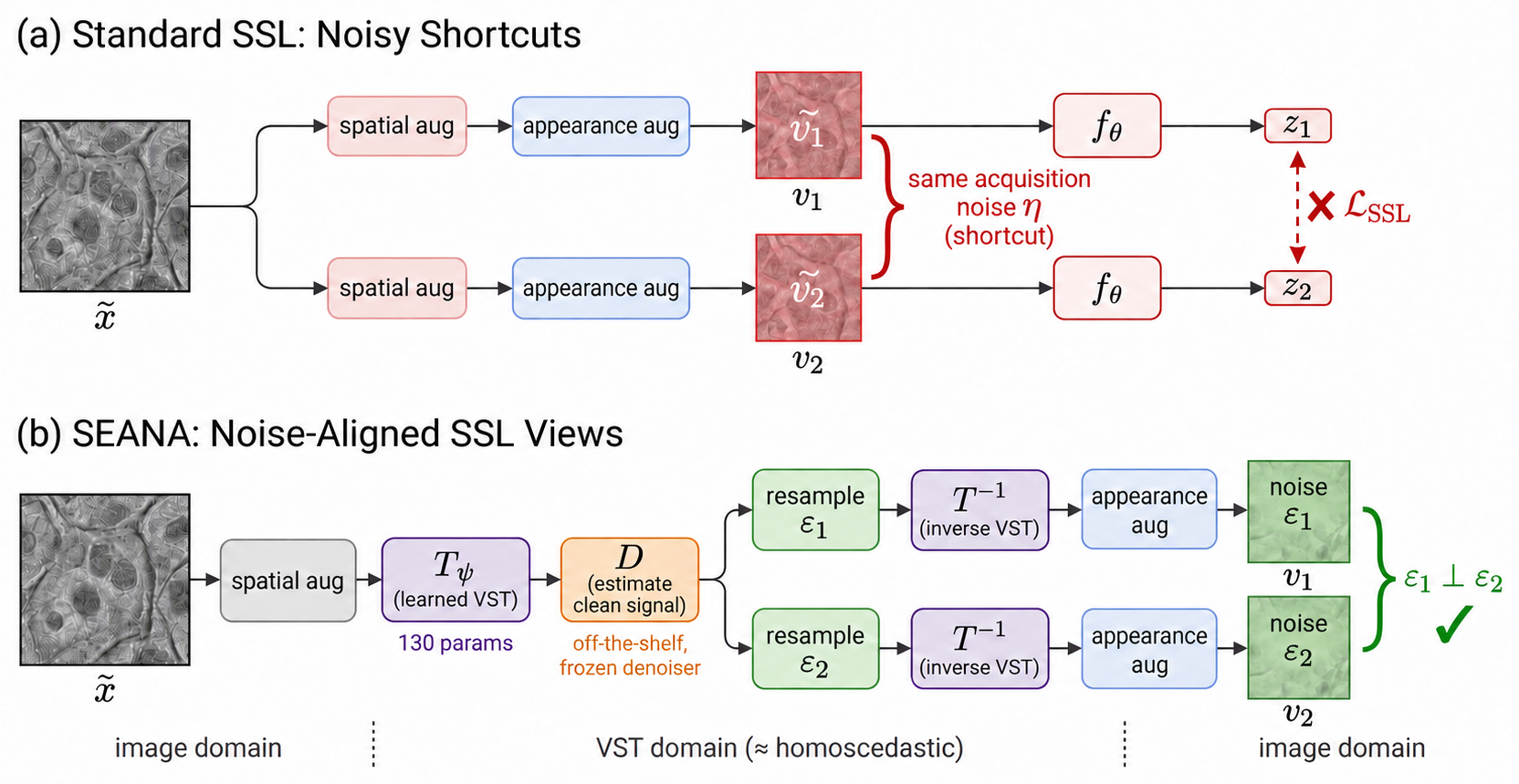}
\caption{\textbf{Overview of SEANA for Noise-Robust Self-Supervised Representation Learning} 
(a)~Standard SSL: both views derive from the same noisy image, creating a shortcut for the encoder to boost agreement by also encoding acquisition noise patterns. 
(b)~SEANA: SEANA learns an invertible variance-stabilizing transform (VST) from noisy data, then estimates the clean signal and resamples independent noise in learned VST space, requiring no clean targets or knowledge of acquistion noise model, no dataset-specific denoiser training or use at inference, and no changes to the SSL objective or encoder.}
\label{fig:overview}
\vspace{-15pt}
\end{figure}

\section{Related Work}
\textbf{SSL under noise:}
Standard joint-embedding SSL methods, such as BYOL
\cite{grill2020bootstrap}, SimCLR \cite{chen2020simple}, and VICReg \cite{bardes2021vicreg}, assume that augmented views differ only in semantically irrelevant factors. \citet{van2025joint} show that when data are corrupted by additive nuisance noise, recovering the noiseless solution requires augmentations that are sufficiently \emph{aligned}
with the corruption. Misalignment cannot be overcome by scaling data alone. They operationalize alignment by appending corruption-matched noise to the augmentation pipeline, observing broad improvements. However, their approach assumes knowledge of the corruption type and
their theoretical analysis focuses on additive noise, leaving signal-dependent corruptions underdeveloped.

\textbf{Self-supervised denoising:}
A natural response to noisy pretraining data is to incorporate denoising. The feasibility of denoiser-guided SSL depends on training denoisers without clean targets. Noise2Noise \cite{lehtinen2018noise2noise} uses pairs of independent noisy observations. Meanwhile, Noise2Self
\cite{batson2019noise2self} and Neighbor2Neighbor (N2N)
\cite{huang2021neighbor2neighbor} construct surrogate supervision from single noisy inputs via blind-spot masking or sub-sampled view pairs. More recently, Next-Scale Prediction (NSP)~\cite{shan2026next}
uses lower-resolution noisy inputs to predict higher-resolution targets,
addressing spatially correlated noise while preserving fine image details. In this work, we use N2N and NSP as baselines (N2N + SSL, NSP + SSL) to test whether direct
denoising of SSL views is sufficient.

\textbf{Denoiser-guided SSL:}
 \citet{lu2025ditch} propose a fully self-supervised paradigm that first trains a self-supervised denoiser on the noisy dataset, then pretrains the SSL backbone with a denoised-to-noisy curriculum and teacher-guided regularization to achieve noise-robust representation learning. While this approach improves over training directly on noisy data, downstream performance on noisy inputs still falls short of applying the denoiser as a preprocessor at both train and test time (i.e. N2N/NSP + SSL). This approach also requires training a dataset-specific denoiser as a first stage and modifies the SSL training procedure. In contrast, SEANA leaves the SSL training curriculum unchanged and does not leverage dataset-specific denoisers.

\textbf{Noise model mismatch and variance stabilization in standard denoising:}
When considering denoisers for use, many off-the-shelf denoisers are trained for additive white Gaussian noise and underperform when the corruption is signal-dependent \cite{plotz2017benchmarking}, necessitating retraining for each dataset. Variance-stabilizing transforms (VSTs) address this by
mapping such heteroscedastic noise to approximately homoscedastic form. Classical VSTs (e.g., Anscombe) require parametric noise models. However, recent data-driven approaches learn the VST nonparametrically.
\citet{wang2021convexvst} formulate VST learning as convex
optimization, and \citet{herbreteau2025self} propose a zero-shot denoising framework that learns an invertible VST using only noisy data and avoids retraining the denoiser.

\textbf{Positioning of SEANA:} SEANA connects noise-alignment in SSL with data-driven variance stabilization. Unlike denoiser-guided SSL
\cite{lu2025ditch}, SEANA does not modify the SSL training procedure or require a dataset-specific denoiser. Unlike \citet{van2025joint}, SEANA investigates and handles signal-dependent noise without assuming knowledge of the corruption model. And unlike VST methods
developed for denoising \cite{wang2021convexvst,herbreteau2025self}, SEANA uses the learned VST not for noise removal but for
\emph{noise-aligned resampling} to enable noise-robust representation learning for improved downstream representation quality.

\section{Noise-Aware View Generation for SSL}
\label{sec:theory}

\subsection{Noise as an Irrelevant Feature in SSL}

Joint-embedding SSL trains an encoder $f_\theta$ to produce similar 
representations for two augmented views $v_1 = \tau_1(\tilde{\mathbf{x}})$, 
$v_2 = \tau_2(\tilde{\mathbf{x}})$ of the same image. Standard 
augmentations (cropping, flipping, color jitter etc.) transform spatial 
and photometric properties. 
Consequently, when both views derive from the same noisy observation 
$\tilde{\mathbf{x}}$, the encoder can increase view agreement by 
encoding shared noise patterns rather than solely semantic content.

\citet{van2025joint} formalize this effect. They model each sample 
as comprising $k$ important signal components and $d - k$ irrelevant 
noise components, with corruption:
\begin{equation}
    \tilde{\mathbf{x}}_i = \mathbf{x}_i + \boldsymbol{\gamma}_i, 
    \quad \boldsymbol{\gamma}_i \sim 
    \mathcal{N}(\mathbf{0}, \boldsymbol{\Gamma}),
    \label{eq:corruption}
\end{equation}
and augmentation distribution 
$\mathcal{T}(\alpha) := \{\tau \mid \tau(\mathbf{x}) = \mathbf{x} 
+ \boldsymbol{\theta} + \alpha\,\boldsymbol{\gamma},\; 
\boldsymbol{\theta} \sim \mathcal{N}(\mathbf{0}, \boldsymbol{\Theta}),\; 
\boldsymbol{\gamma} \sim \mathcal{N}(\mathbf{0}, \boldsymbol{\Gamma})\}$, 
where $\boldsymbol{\theta}$ captures standard augmentation variability 
and the term $\alpha\,\boldsymbol{\gamma}$, drawn from the 
\emph{same distribution} as the corruption, injects noise-aligned 
variability. Their key result establishes a minimum alignment 
threshold for joint-embedding SSL.

\textbf{The Alignment Condition: } The joint-embedding SSL model trained on corrupted data 
$\tilde{\mathbf{X}}$ recovers the clean-data solution 
$\mathbf{W}^\star$ as $n \to \infty$ if and only if
\begin{equation}
    \alpha^2 > \alpha^2_{\mathrm{JE}} := 
    \max_{i \in \llbracket k{+}1 : d \rrbracket} \,
    \frac{1 - \delta}{\delta} 
    - \frac{\lambda_i^{\boldsymbol{\Theta}}}
           {\lambda_i^{\boldsymbol{\Gamma}}}\,,
    \label{eq:alignment}
\end{equation}
where $\lambda_i^{\boldsymbol{\Gamma}}$ and 
$\lambda_i^{\boldsymbol{\Theta}}$ are the eigenvalues of the noise 
and augmentation covariances on the noise components, and 
$\delta = \min_{i \in \llbracket k \rrbracket} 
\frac{\frac{1}{n}\kappa_i^2}{\frac{1}{n}\kappa_i^2 + 
\lambda_i^{\boldsymbol{\Theta}}}$ depends on the signal-to-noise 
ratio.

This result has two important implications. First, unlike supervised 
learning, increasing the dataset size $n$ alone cannot overcome 
misalignment, the threshold $\alpha^2_{\mathrm{JE}}$ persists in 
the infinite-sample limit. Second, when the alignment condition 
\emph{is} satisfied, the noise-aligned augmentation renders the 
noise across views conditionally independent given the clean signal, 
preventing the encoder from exploiting shared noise patterns.

\subsection{Signal-Dependent Noise and Variance Stabilization}

The alignment condition of Equation~\eqref{eq:alignment} requires 
augmenting with noise drawn from 
$\mathcal{N}(\mathbf{0}, \boldsymbol{\Gamma})$. For the 
homoscedastic (constant variance) case analyzed by \citet{van2025joint}, 
$\boldsymbol{\Gamma}$ is constant and could in principle be 
estimated. In scientific imaging, however, corruption is typically 
\emph{signal-dependent} \cite{foi2009clipped}. Specifically, a convenient local 
approximation for pixelwise independent, zero-mean noise is
\begin{equation}
    \tilde{x}_p = x_p + \sigma(x_p)\,\varepsilon_p, \quad 
    \varepsilon_p \overset{\mathrm{i.i.d.}}{\sim} 
    \mathcal{N}(0, 1),
    \label{eq:signalnoise}
\end{equation}
where $p$ indexes pixels and $\sigma(\cdot)$ is an unknown, 
signal-dependent scale. 
The effective noise covariance $\boldsymbol{\Gamma}$ now has entries 
$\sigma^2(x_p)$ that vary with the clean signal, making direct 
resampling from $\mathcal{N}(\mathbf{0}, \boldsymbol{\Gamma})$ 
intractable without knowing $\sigma(\cdot)$.

A \emph{variance-stabilizing transform} (VST) reduces this 
signal-dependent case to the homoscedastic case where the alignment 
condition can be directly satisfied. A VST is a monotone mapping 
$T : \mathbb{R} \to \mathbb{R}$ that compresses intensity regions 
where noise variance is high and stretches regions where it is low, 
so that after transformation the residual corruption becomes 
approximately homoscedastic:
\begin{equation}
    T(\tilde{x}_p) \approx T(x_p) + \eta_p, \quad 
    \eta_p \overset{\mathrm{i.i.d.}}{\sim} 
    \mathcal{N}(0, \sigma_T^2),
    \label{eq:vst}
\end{equation}
with roughly constant $\sigma_T^2$ across intensities. In the 
transformed domain, $\boldsymbol{\Gamma} \approx 
\sigma_T^2\,\mathbf{I}$, and the alignment condition of 
Equation~\eqref{eq:alignment} can be satisfied by resampling 
i.i.d.\ Gaussian noise with constant variance, requiring no 
knowledge of $\sigma(\cdot)$. Classical analytic VSTs (e.g., 
Anscombe-type transforms) achieve Equation~\eqref{eq:vst} under 
specific parametric assumptions \cite{anscombe1948transformation, bartlett1947use}. 
SEANA instead learns $T$ directly from the noisy dataset, 
enabling noise-aligned view generation without specifying the 
noise model or parameters.

\section{Method}

Given the framework of Section~\ref{sec:theory}, SEANA's design 
follows directly. We need to: (a)~learn an invertible VST $T$ from 
noisy data without clean targets or noise model knowledge, and 
(b)~use $T$ to generate noise-aligned views for SSL. SEANA leaves 
the SSL objective and encoder architecture unchanged.

\subsection{Learning an Invertible Spline VST from Noisy Data}

SEANA learns an invertible VST $T_\psi$ and its inverse 
$T_\psi^{-1}$ from the noisy training split, using a frozen 
off-the-shelf Gaussian denoiser as a calibration tool.

\paragraph{Spline parameterization:}
We parameterize $T_\psi$ as a strictly increasing, piecewise-linear 
spline over a bounded intensity range. Let $\{u_k\}_{k=1}^K$ be 
fixed knot locations on $[0,1]$ and $\{y_k\}_{k=1}^K$ the learnable 
knot values. $T_\psi$ is defined by linear interpolation between 
adjacent knots. To guarantee invertibility, we enforce strict 
monotonicity by parameterizing knot increments with an exponential 
mapping (details in Appendix~\ref{app:cpwl_vst}). We also learn a 
lightweight affine correction in the inverse map to absorb small 
systematic biases:
\begin{equation}
    T_\psi^{-1}(t) \leftarrow T_\psi^{-1}(t) + \alpha\, t + \beta,
\end{equation}
with learnable scalars $(\alpha, \beta)$. The total number of 
trainable parameters is $K + 2$ (128 knot values plus two affine 
terms in our experiments).

\paragraph{Blind-spot self-supervision with a frozen denoiser:}
Let $D_{\sigma_d}$ be a pretrained Gaussian denoiser with noise 
level parameter $\sigma_d$. We use $D_{\sigma_d}$ as a fixed 
operator and train only the VST parameters $\psi$. Since no clean 
targets are available, we adopt a blind-spot (J-invariant) strategy 
that predicts masked pixels from their neighborhood.

For a patch $\mathbf{z}$ (normalized to $[0,1]$), we:
(i)~transform to VST space: $\mathbf{t} = T_\psi(\mathbf{z})$
(ii)~sample a binary mask $\mathbf{M}$ and construct a blind-spot 
input by replacing masked pixels with a local average (Donut 
filter):
\begin{equation}
    \mathbf{t}_{\mathrm{bs}} = (1 - \mathbf{M}) \odot \mathbf{t} 
    + \mathbf{M} \odot \mathrm{Donut}(\mathbf{t})
\end{equation}
(iii)~denoise in VST space and invert: 
$\widehat{\mathbf{z}} = T_\psi^{-1}(D_{\sigma_d}
(\mathbf{t}_{\mathrm{bs}}))$.
We minimize the MSE on masked pixels only:
\begin{equation}
    \mathcal{L}_{\mathrm{VST}}(\psi) = 
    \left\| \mathbf{M} \odot 
    \left(\widehat{\mathbf{z}} - \mathbf{z}\right) 
    \right\|_2^2.
    \label{eq:vst_loss}
\end{equation}
Because the masked pixels are excluded from the denoiser input, the 
prediction at those locations cannot trivially copy the observation. Under the pixelwise independence assumption of 
Equation~\eqref{eq:signalnoise}, this yields a self-supervised 
learning signal for $\psi$. We optimize 
Equation~\eqref{eq:vst_loss} over random patches from the training 
split with random geometric augmentations.

\subsection{SEANA View Generation for SSL}
\label{sec:seana_view}

After calibration, SEANA replaces the standard view generator with 
a composition of standard spatial augmentations and VST-space 
denoising with resampling. Given a noisy image $\tilde{\mathbf{x}}$ 
and random augmentation parameters $\omega$:
\begin{align}
    \mathbf{z} &= \mathcal{S}(\tilde{\mathbf{x}};\,\omega), 
    \label{eq:seana_start} \\
    \mathbf{t} &= T_\psi(\mathbf{z}), \\
    \widehat{\mathbf{t}} &= D_{\sigma_d}(\mathbf{t}), \\
    \mathbf{t}_{\mathrm{aug}} &= \widehat{\mathbf{t}} 
    + \sigma_{\mathrm{vst}}\,\boldsymbol{\varepsilon}, 
    \quad \boldsymbol{\varepsilon} \sim 
    \mathcal{N}(\mathbf{0}, \mathbf{I}), \\
    \mathbf{v} &= T_\psi^{-1}(\mathbf{t}_{\mathrm{aug}}),
    \label{eq:seana_end}
\end{align}
where $\mathcal{S}$ denotes standard spatial augmentations (random 
crop, flip) and $\sigma_{\mathrm{vst}} \geq 0$ controls the 
resampling strength. Two views are formed using independent spatial 
parameters $\omega_1, \omega_2$ and independent noise draws 
$\boldsymbol{\varepsilon}_1, \boldsymbol{\varepsilon}_2$. Each view can then be given standard appearance augmentations and then 
passed to any joint-embedding SSL objective without modification.

When $\sigma_{\mathrm{vst}} > 0$, SEANA resamples i.i.d.\ Gaussian 
perturbations in the stabilized domain, producing independent noise 
realizations that approximately satisfy the alignment condition of 
Equation~\eqref{eq:alignment}. Setting $\sigma_{\mathrm{vst}} = 0$ 
reduces SEANA to VST denoising, which can eliminate the 
stochastic view diversity that SSL objectives tend to benefit from 
(Section~\ref{sec:ablation}).

\subsection{Practical Considerations}

\paragraph{Hyperparameters:}
SEANA exposes two noise-related hyperparameters: $\sigma_d$ (the 
noise level provided to the frozen denoiser) and 
$\sigma_{\mathrm{vst}}$ (the resampling strength in VST space). In 
practice, $\sigma_{\mathrm{vst}}$ can be tuned on a held-out split 
or estimated from the empirical residual standard deviation 
$\|\widehat{\mathbf{t}} - \mathbf{t}\|$.

\paragraph{Computational cost:}
VST calibration is performed once per training split via patch-based optimization of Equation~\eqref{eq:vst_loss}. In the main paper, we fit 130 parameters over 5{,}000 steps. To eliminate computational bottlenecks during SSL training, we first cache the fitted VST + denoiser's clean signal estimate for the full images offline. During training, we simply apply spatial augmentations to the cached images and perform VST-space noise resampling, effectively reducing the denoiser overhead to a one-time preprocessing step done before SSL training. At downstream evaluation, the encoder is used \emph{without} any denoising or VST module.

\section{Experiments \& Results}
\label{sec:experiments}

We evaluate SEANA across five experiments: (i) representation invariance analysis to isolate  noise shortcuts, (ii)~residual analysis 
validating the VST mechanism, (iii)~synthetic noise models on CIFAR-10 and 
ImageNet-100 for SSL representation quality, (iv)~real fluorescence imaging scenario, (v)~ablation of SEANA in real fluorescence imaging. We define the synthetic noise models (Gaussian, Poisson-Gaussian, Multiplicative) used in all experiments in Appendix \ref{app:noise_models}. 

We also experiment with a real clinical ultrasound imaging dataset in Appendix \ref{app:fetal_exp}. For all experiments, SEANA leverages DRUNet \cite{zhang2021plug} for its blind-spot frozen, off-the-shelf gaussian denoiser, but we provide an ablation with others in Appendix \ref{app:den_abl}.
Further implementation/training details are provided in 
Appendix ~\ref{app:impl_details}, spline knot count sensitivity in Appendix ~\ref{app:knot_sweep}, and hyperparameter sensitivity in Appendix ~\ref{app:sigmasense}.

\begin{figure}[t]
\centering
\begin{subfigure}[t]{0.5\textwidth}
    \centering
    \includegraphics[width=\linewidth, height=5cm, keepaspectratio]{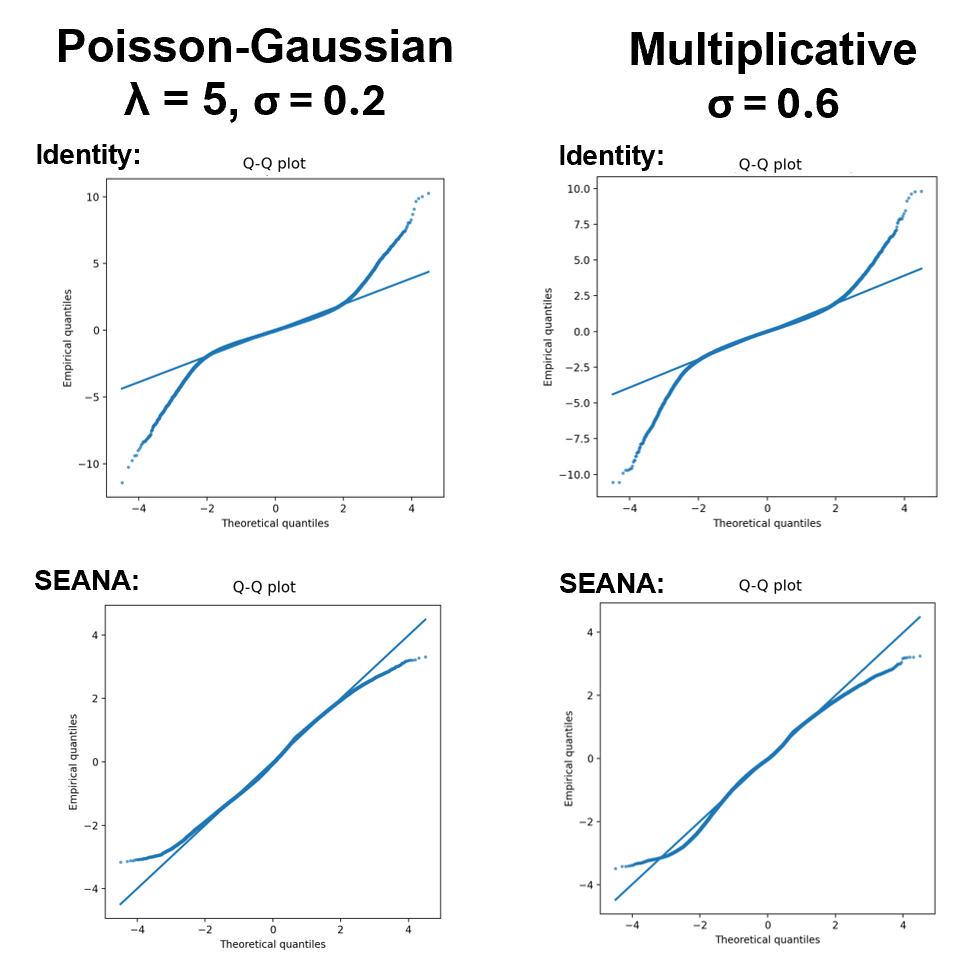}
    \caption{CIFAR-10}
    \label{fig:qq_cifar10}
\end{subfigure}
\hfill 
\begin{subfigure}[t]{0.45\textwidth}
    \centering
    \includegraphics[width=\linewidth, height=5cm, keepaspectratio]{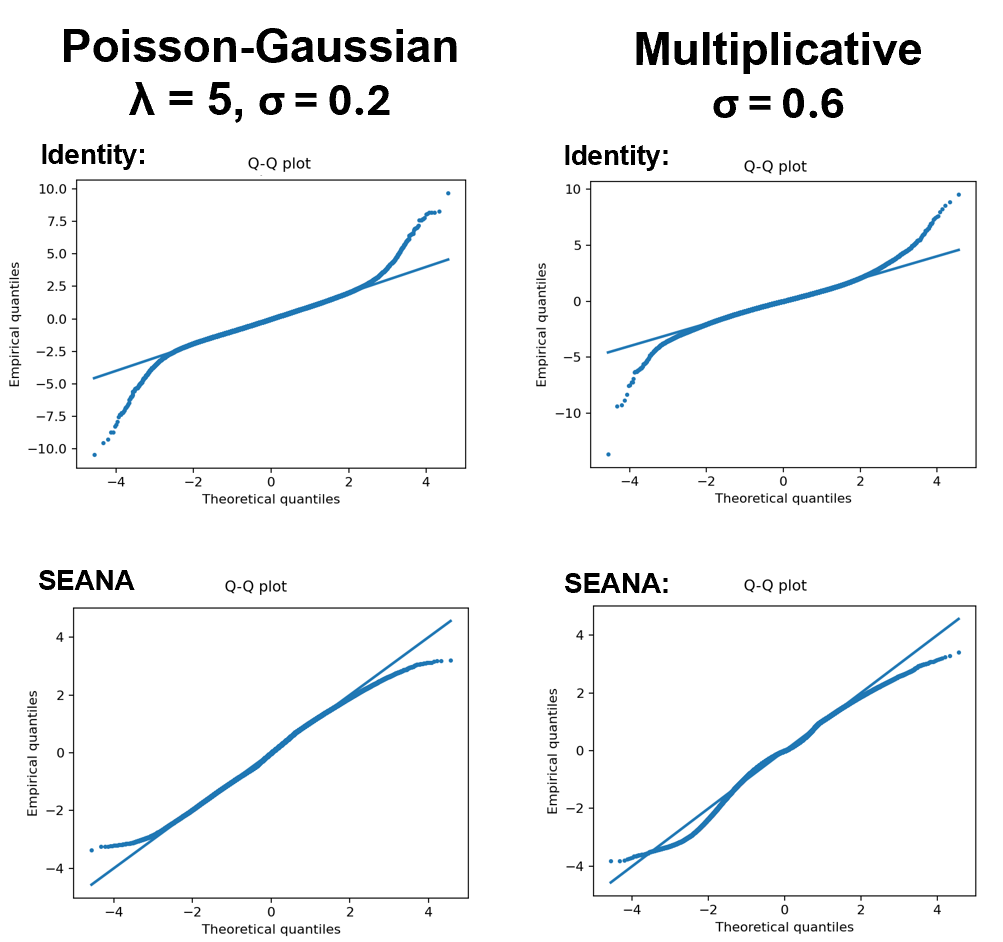}
    \caption{ImageNet-100}
    \label{fig:qq_imagenet100}
\end{subfigure}
\caption{\textbf{Q-Q plots.} Residual quantiles vs.\ standard normal for identity and SEANA in the highest noise regimes of Poisson-Gaussian/Multiplicative (Speckle). A straight diagonal indicates Gaussianity. SEANA consistently produces the straightest Q-Q plots.}
\label{fig:qq_plots}
\vspace{-12pt}
\end{figure}

\subsection{Learned Representation Noise-Invariance Analysis}

We study whether SEANA helps SSL models learn representations that are less
sensitive to noise while preserving image content, isolating the noise shortcut mechanism. We compare backbones trained
with baseline SSL and SEANA under synthetic noise models applied to CIFAR-10.

\paragraph{Setup:} For each clean test image, we generate $K=8$ independent noise realizations and
pass the resulting noisy images directly through the frozen backbone. We then
group the embeddings by their original clean-image identity and use a one-way
ANOVA decomposition to measure two sources of variation: changes within the
same image as the noise varies, and differences between images. We evaluate under six separate synthetic noise conditions: additive Gaussian ($\sigma \in \{0.1, 0.5\}$), Poisson--Gaussian ($\lambda \in \{30, 5\}$), and multiplicative speckle ($\sigma \in \{0.2, 0.6\}$). Each condition spans a low-noise and high-noise regime. Our baseline SSL approaches are BYOL, SimCLR, and VICReg.

\paragraph{Noise-driven variance fraction:}
For each image, we average the embeddings of its eight noisy versions.
We measure $S_W$ as the mean within-image variance (the 8 embeddings of the same clean image spread apart only because the noise differs), i.e noise-driven variance. We measure $S_B$ as the variation
between the per-image averages, i.e content.

Specifically, the within-image and between-image dispersions are:
\begin{equation}
\begin{aligned}
    S_W &= \frac{1}{NK}\sum_{i=1}^{N}\sum_{r=1}^{K}
             \lVert z_{ir}-\bar z_i\rVert_2^2,\\
    S_B &= \frac{1}{N}\sum_{i=1}^{N}
             \lVert\bar z_i-\bar z\rVert_2^2.
\end{aligned}
\end{equation}
Where $z_\textit{ir}$ is the embedding of image i under its r-th independent noise realization, $\bar{z}_i$ is the average embedding of the K noisy versions of image i, and $\bar{z}$ is the overall average embedding across all images and noise realizations. Our primary metric is then:
\begin{equation}
    \nu = \frac{S_W}{S_W+S_B} = 1-\eta^2,
    \qquad
    \eta^2 = \frac{S_B}{S_W+S_B}.
    \label{eq:noise_variance_fraction}
\end{equation}
Lower $\nu$ indicates the representation barely moves when only the noise changes, meaning it's more noise-invariant.

\begin{table}[t]
    \centering
    \small
    \setlength{\tabcolsep}{5pt}
    \caption{Noise-driven variance fraction $\nu=1-\eta^2$ under six
    synthetic noise conditions applied to CIFAR-10. Lower is better. Entries are arithmetic means across BYOL, SimCLR, VICReg. PG denotes Poisson--Gaussian noise.}
    \label{tab:representation_noise_fraction}
    \begin{tabular}{lrrr}
        \toprule
        Noise condition & Baseline & SEANA & Reduction \\
        \midrule
        Gaussian (low)  & 0.021 & 0.013 & 40.3\% \\
        Gaussian (high) & 0.115 & 0.063 & 45.1\% \\
        PG (low)        & 0.026 & 0.019 & 25.5\% \\
        PG (high)       & 0.079 & 0.051 & 35.8\% \\
        Speckle (low)   & 0.020 & 0.016 & 23.3\% \\
        Speckle (high)  & 0.051 & 0.036 & 29.9\% \\
        \bottomrule
    \end{tabular}
\end{table}

\paragraph{Results across noise conditions:}
SEANA lowers the mean noise-driven variance fraction in all six conditions
(Table~\ref{tab:representation_noise_fraction}), with relative reductions of
$23.3$--$45.1\%$. In the high-noise conditions, the baseline fraction ranges
from $5.1\%$ to $11.5\%$, and SEANA reduces it by $29.9$--$45.1\%$ relative
to baseline.

\subsection{Residuals in VST Space}
\label{sec:exp_residuals}

\paragraph{Setup:}
SEANA assumes that its learned VST maps signal-dependent noise to 
approximately homoscedastic Gaussian residuals. We test this on 
CIFAR-10 and ImageNet-100 under four synthetic non-Gaussian corruptions 
(low/high Poisson--Gaussian and low/high speckle). For each 
corrupted image $\tilde{\mathbf{x}}$, a candidate transform $T$ 
produces $\mathbf{y} = T(\tilde{\mathbf{x}})$, a denoised estimate 
$\mathbf{s} = D(\mathbf{y})$, and residual 
$\mathbf{r} = \mathbf{y} - \mathbf{s}$. We compare three 
transforms: the identity (no stabilization), the classical 
analytic VST (generalized Anscombe for Poisson--Gaussian, log for 
speckle), and SEANA's learned VST.

\paragraph{Metrics:}
We evaluate residuals on three axes: (i)~\emph{variance stability} 
(coefficient of variation of residual variance across intensity 
bins, lower indicates more homoscedastic residuals), 
(ii)~\emph{Gaussianity} (skewness and excess kurtosis of 
standardized residuals, closer to zero is better), and 
(iii)~Q-Q plots against a standard normal distribution (a straight diagonal indicates Gaussianity).

\paragraph{Results:}
Figure~\ref{fig:qq_cifar10} and \ref{fig:qq_imagenet100} shows qualitative Q-Q plots 
for the high-noise regimes, where SEANA's residuals follow a straighter diagonal (more Gaussian). Table~\ref{tab:vst_residuals} reports quantitative metrics on CIFAR-10 \& ImageNet-100. SEANA achieves the lowest variance stabilization (Var CV) in 3 of 4 
conditions and the best excess kurtosis in 3 of 4 conditions, indicating that the learned VST produces more homoscedastic and more Gaussian residuals than both the identity and the classical VST. Notably, the classical log transform \emph{worsens} residual quality for speckle noise (kurtosis 2.45 and 2.57 vs.\ identity's 
0.93 and 0.29), likely due to boundary effects from intensity clamping. SEANA avoids this failure mode by adapting to the actual data distribution.

\begin{table}[t]
\caption{\textbf{Residual analysis:} For each transform $T$, we 
compute residuals $\mathbf{r} = T(\tilde{\mathbf{x}}) - 
D(T(\tilde{\mathbf{x}}))$ and measure variance stabilization 
(Var CV $\downarrow$), skewness ($|\cdot|\downarrow$), and excess 
kurtosis ($|\cdot|\downarrow$).}
\label{tab:vst_residuals}
\centering
\footnotesize
\setlength{\tabcolsep}{1.4pt}

\begin{minipage}[t]{0.48\textwidth}
\centering
\textbf{CIFAR-10}\\[4pt]
\begin{tabular}{llccc}
\toprule
\textbf{Noise} & \textbf{Transform} & 
\textbf{Var CV}$\downarrow$ & \textbf{Skew.}$\downarrow$ & \textbf{Kurt.}$\downarrow$ \\
\midrule
\multirow{3}{*}{\shortstack[l]{Poisson-Gaussian\\[-1pt]\scriptsize$\lambda{=}30,\sigma_g{=}.01$}}
& Identity  & 0.400 & 0.054  & 0.180 \\
& Classical & 0.279 & 0.211 & 0.078 \\
& SEANA     & 0.215 & 0.017 & 0.049 \\
\midrule
\multirow{3}{*}{\shortstack[l]{Poisson-Gaussian\\[-1pt]\scriptsize$\lambda{=}5,\sigma_g{=}.2$}}
& Identity  & 0.544 & 0.138  & 0.344 \\
& Classical & 0.498 & 0.134 & 0.372 \\
& SEANA     & 0.356 & 0.101 & 0.567 \\
\midrule
\multirow{3}{*}{\shortstack[l]{Multiplicative\\[-1pt]\scriptsize$\sigma{=}0.2$}}
& Identity  & 0.654 & 0.157 & 0.931 \\
& Classical & 0.793 & 0.707 & 2.445 \\
& SEANA     & 0.411 & 0.512 & 0.805 \\
\midrule
\multirow{3}{*}{\shortstack[l]{Multiplicative\\[-1pt]\scriptsize$\sigma{=}0.6$}}
& Identity  & 0.561 & 0.006 & 0.286 \\
& Classical & 1.110 & 0.305 & 2.572 \\
& SEANA     & 0.707 & 0.396 & 0.040 \\
\bottomrule
\end{tabular}
\end{minipage}%
\hfill
\begin{minipage}[t]{0.48\textwidth}
\centering
\textbf{ImageNet-100}\\[4pt]
\begin{tabular}{llccc}
\toprule
\textbf{Noise} & \textbf{Transform} & 
\textbf{Var CV}$\downarrow$ & \textbf{Skew. $\downarrow$} & \textbf{Kurt. $\downarrow$} \\
\midrule
\multirow{3}{*}{\shortstack[l]{Poisson-Gaussian\\[-1pt]\scriptsize$\lambda{=}30,\sigma_g{=}.01$}}
& Identity  & 0.006 & 0.092 & 0.598 \\
& Classical & 0.002 & 0.253 & 0.124 \\
& SEANA     & 0.001 & 0.031 & 0.097 \\
\midrule
\multirow{3}{*}{\shortstack[l]{Poisson-Gaussian\\[-1pt]\scriptsize$\lambda{=}5,\sigma_g{=}.2$}}
& Identity  & 0.004 & 0.121 & 2.010 \\
& Classical & 0.012 & 0.103 & 0.422 \\
& SEANA     & 0.001 & 0.039 & 0.547 \\
\midrule
\multirow{3}{*}{\shortstack[l]{Multiplicative\\[-1pt]\scriptsize$\sigma{=}0.2$}}
& Identity  & 0.002 & 0.178 & 0.830 \\
& Classical & 0.037 & 0.751 & 2.035 \\
& SEANA     & 0.001 & 0.044 & 0.817 \\
\midrule
\multirow{3}{*}{\shortstack[l]{Multiplicative\\[-1pt]\scriptsize$\sigma{=}0.6$}}
& Identity  & 0.003 & 0.215 & 0.282 \\
& Classical & 0.046 & 0.098 & 2.147 \\
& SEANA     & 0.002 & 0.064 & 0.477 \\
\bottomrule
\end{tabular}
\end{minipage}

\end{table}

\subsection{SSL Representation Quality under Synthetic Noise on CIFAR-10 and ImageNet-100}
\label{sec:exp_synthetic}

\paragraph{Setup:}
We evaluate SEANA on CIFAR-10 \cite{krizhevsky2009learning} and 
ImageNet-100 \cite{deng2009imagenet} under six separate synthetic noise conditions: additive Gaussian ($\sigma \in \{0.1, 0.5\}$), Poisson--Gaussian ($\lambda \in \{30, 5\}$), and multiplicative speckle ($\sigma \in \{0.2, 0.6\}$). Each condition spans a low-noise and high-noise regime. We train three joint-embedding SSL methods (BYOL, SimCLR, and VICReg) with a ResNet-50 encoder on noisy training data and evaluate via linear probing top-1 classification accuracy on the \emph{clean} test set, as a way to measure learned representation quality. All results report mean $\pm$ std over 3 seeds. We compare against two baselines: standard SSL trained on the noisy images (Base) and a pipeline training Neighbor2Neighbor (N2N) \cite{huang2021neighbor2neighbor} 
denoiser on the noisy data followed by each SSL model (BYOL, SimCLR, and VICReg). We also include a more granular analysis on the synthetic CIFAR-10 scenarios using the NSP denoiser, N2N / NSP denoisers + resampling, and analytical VST in Appendix \ref{app:oracle_comparison}

\paragraph{Results:}
Tables~\ref{tab:cifar10_split} and \ref{tab:imagenet100_split} summarize the results. SEANA improves over both baselines across all 36 conditions (6 noise types $\times$ 3 SSL methods $\times$ 
2 datasets). On CIFAR-10, gains over the noisy baseline range from +6.4~pp (VICReg, Gaussian $\sigma{=}0.1$) to +20.5~pp (VICReg, PG $\lambda{=}5$), with the largest improvements consistently occurring under severe corruption. On ImageNet-100, SEANA improves across all conditions, with gains reaching +8.4~pp (BYOL, Shot $\lambda{=}5$) and +7.0~pp (VICReg, Shot $\lambda{=}30$).

Additionally, two patterns emerge. SEANA exhibits markedly lower variance across seeds, for example, on CIFAR-10 at PG $\lambda{=}5$, VICReg baseline variance is $\pm$2.1 while SEANA achieves $\pm$0.3, suggesting more stable optimization under noise. Second, N2N denoising frequently \emph{hurts} linear-probing top-1 performance on clean-test data relative to the noisy baseline, indicating a decrease in representation quality. On CIFAR-10 at PG $\lambda{=}5$, N2N degrades SimCLR from 31.8 to 24.5 and VICReg from 34.2 to 24.9. On ImageNet-100 at Gaussian $\sigma{=}0.5$, N2N drops VICReg from 39.6 to 35.5. This points that naive denoising can perhaps remove the stochastic view diversity that SSL objectives require.

\begin{table}[t]
\caption{\textbf{CIFAR-10 linear probing accuracy (\%).} 
Clean-test top-1 accuracy under synthetic noisy training data to measure representation quality. SEANA is our method. N2N = Neighbor2Neighor Denoiser + SSL Baseline. PG = Poisson-Gaussian Noise. Gauss = Additive Gaussian Noise. Speckle = Speckle (Multiplicative) Noise. Best in \textbf{bold}.}
\label{tab:cifar10_split}
\centering
\footnotesize 
\setlength{\tabcolsep}{1.2pt} 
\resizebox{\textwidth}{!}{ 
\begin{tabular}{llcccc @{\hskip 8pt} llcccc} 
\toprule
\textbf{Noise} & \textbf{SSL} & \textbf{Base} & \textbf{N2N} & \textbf{SEANA} & $\boldsymbol{\Delta}$ &
\textbf{Noise} & \textbf{SSL} & \textbf{Base} & \textbf{N2N} & \textbf{SEANA} & $\boldsymbol{\Delta}$ \\
\midrule

\multirow{3}{*}{\shortstack[l]{(Low) Gauss\\ \scriptsize$\sigma{=}0.1$}}
& BYOL   & 52.6\pms{1.1} & 53.1\pms{0.7} & \textbf{60.8\pms{0.2}} & \up{8.2} &
\multirow{3}{*}{\shortstack[l]{(High) PG\\ \scriptsize$\lambda{=}5, \sigma_g{.2}$}}
& BYOL   & 30.0\pms{0.9} & 26.9\pms{1.8} & \textbf{47.8\pms{0.6}} & \up{17.8} \\
& SimCLR & 59.4\pms{0.7} & 59.6\pms{0.7} & \textbf{66.0\pms{0.2}} & \up{6.6} &
& SimCLR & 31.8\pms{2.1} & 24.5\pms{0.5} & \textbf{50.5\pms{0.4}} & \up{18.7} \\
& VICReg & 63.0\pms{0.4} & 64.2\pms{0.7} & \textbf{69.4\pms{0.1}} & \up{6.4} &
& VICReg & 34.2\pms{2.1} & 24.9\pms{2.4} & \textbf{54.7\pms{0.3}} & \up{20.5} \\

\midrule

\multirow{3}{*}{\shortstack[l]{(High) Gauss\\ \scriptsize$\sigma{=}0.5$}}
& BYOL   & 28.4\pms{0.2} & 25.3\pms{0.7} & \textbf{41.9\pms{0.4}} & \up{13.5} &
\multirow{3}{*}{\shortstack[l]{(Low) Speckle\\ \scriptsize$\sigma{=}0.2$}}
& BYOL   & 51.3\pms{1.2} & 50.6\pms{0.2} & \textbf{61.5\pms{0.6}} & \up{10.3} \\
& SimCLR & 27.5\pms{0.9} & 26.8\pms{0.4} & \textbf{42.2\pms{0.5}} & \up{14.7} &
& SimCLR & 58.4\pms{0.2} & 58.2\pms{0.5} & \textbf{67.0\pms{0.1}} & \up{8.6} \\
& VICReg & 29.8\pms{1.1} & 30.1\pms{1.9} & \textbf{47.4\pms{0.6}} & \up{17.6} &
& VICReg & 63.6\pms{0.5} & 63.4\pms{0.1} & \textbf{71.4\pms{0.4}} & \up{7.8} \\

\midrule

\multirow{3}{*}{\shortstack[l]{(Low) PG\\ \scriptsize$\lambda{=}30, \sigma_g{.01}$}}
& BYOL   & 47.2\pms{0.6} & 46.2\pms{1.3} & \textbf{59.9\pms{0.3}} & \up{12.7} &
\multirow{3}{*}{\shortstack[l]{(High) Speckle\\ \scriptsize$\sigma{=}0.6$}}
& BYOL   & 33.3\pms{1.4} & 33.6\pms{0.3} & \textbf{50.1\pms{0.3}} & \up{16.8} \\
& SimCLR & 54.1\pms{0.5} & 54.3\pms{1.1} & \textbf{65.0\pms{0.0}} & \up{10.8} &
& SimCLR & 36.7\pms{0.6} & 37.7\pms{1.2} & \textbf{54.7\pms{0.5}} & \up{18.0} \\
& VICReg & 57.8\pms{1.1} & 59.3\pms{1.3} & \textbf{69.1\pms{0.1}} & \up{11.3} &
& VICReg & 41.4\pms{1.8} & 42.4\pms{3.4} & \textbf{58.9\pms{0.4}} & \up{17.5} \\

\bottomrule
\end{tabular}
}
\end{table}

\begin{table}[t]
\caption{\textbf{ImageNet-100 linear probing accuracy (\%):} 
Clean-test top-1 accuracy under synthetic noisy training data to measure representation quality. SEANA is our method. N2N = Neighbor2Neighbor Denoiser + SSL Baseline. Best in \textbf{bold}.}
\label{tab:imagenet100_split}
\centering
\footnotesize 
\setlength{\tabcolsep}{1.2pt} 
\resizebox{\textwidth}{!}{ 
\begin{tabular}{llcccc @{\hskip 8pt} llcccc} 
\toprule
\textbf{Noise} & \textbf{SSL} & \textbf{Base} & \textbf{N2N} & \textbf{SEANA} & $\boldsymbol{\Delta}$ &
\textbf{Noise} & \textbf{SSL} & \textbf{Base} & \textbf{N2N} & \textbf{SEANA} & $\boldsymbol{\Delta}$ \\
\midrule

\multirow{3}{*}{\shortstack[l]{(Low) Gauss\\ \scriptsize$\sigma{=}0.1$}}
& BYOL   & 55.9\pms{0.9} & 57.5\pms{0.5} & \textbf{60.5\pms{0.2}} & \up{4.6} &
\multirow{3}{*}{\shortstack[l]{(High) Shot\\ \scriptsize$\lambda{=}5$}}
& BYOL   & 42.6\pms{0.3} & 43.5\pms{1.0} & \textbf{50.9\pms{0.3}} & \up{8.4} \\
& SimCLR & 51.9\pms{0.3} & 53.7\pms{0.2} & \textbf{58.0\pms{0.1}} & \up{6.1} &
& SimCLR & 40.8\pms{0.9} & 41.1\pms{0.0} & \textbf{47.1\pms{0.1}} & \up{6.3} \\
& VICReg & 54.1\pms{0.2} & 55.1\pms{0.3} & \textbf{59.3\pms{0.6}} & \up{5.3} &
& VICReg & 42.5\pms{0.4} & 41.0\pms{0.5} & \textbf{49.1\pms{0.2}} & \up{6.6} \\

\midrule

\multirow{3}{*}{\shortstack[l]{(High) Gauss\\ \scriptsize$\sigma{=}0.5$}}
& BYOL   & 39.2\pms{0.7} & 38.0\pms{0.3} & \textbf{46.5\pms{0.4}} & \up{7.2} &
\multirow{3}{*}{\shortstack[l]{(Low) Speckle\\ \scriptsize$\sigma{=}0.2$}}
& BYOL   & 54.8\pms{0.4} & 57.3\pms{0.8} & \textbf{59.8\pms{0.3}} & \up{4.9} \\
& SimCLR & 38.7\pms{0.8} & 36.9\pms{0.4} & \textbf{44.5\pms{0.4}} & \up{5.7} &
& SimCLR & 51.0\pms{0.5} & 53.9\pms{0.3} & \textbf{56.5\pms{0.6}} & \up{5.5} \\
& VICReg & 39.6\pms{0.7} & 35.5\pms{0.5}& \textbf{45.4\pms{0.2}} & \up{5.8} &
& VICReg & 51.6\pms{0.8} & 55.1\pms{0.9} & \textbf{58.3\pms{0.9}} & \up{6.7} \\

\midrule

\multirow{3}{*}{\shortstack[l]{(Low) Shot\\ \scriptsize$\lambda{=}30$}}
& BYOL   & 53.3\pms{0.7} & 56.0\pms{0.5} & \textbf{59.3\pms{0.3}} & \up{6.0} &
\multirow{3}{*}{\shortstack[l]{(High) Speckle\\ \scriptsize$\sigma{=}0.6$}}
& BYOL   & 47.5\pms{0.0} & 48.5\pms{0.7} & \textbf{49.3\pms{0.4}} & \up{1.8} \\
& SimCLR & 48.8\pms{0.3} & 52.2\pms{0.5} & \textbf{55.2\pms{0.4}} & \up{6.4} &
& SimCLR & 42.5\pms{0.1} & 44.1\pms{0.3} & \textbf{45.4\pms{0.3}} & \up{3.0} \\
& VICReg & 50.3\pms{0.3} & 54.2\pms{0.7} & \textbf{57.3\pms{0.8}} & \up{7.0} &
& VICReg & 43.8\pms{0.7} & 43.8\pms{0.5} & \textbf{46.6\pms{0.2}} & \up{2.8} \\

\bottomrule
\end{tabular}
}
\end{table}

\subsection{SSL on Real-World Single-Cell Fluorescence Imaging with Signal-Dependent Noise}
\label{sec:exp_bbbc048}

\paragraph{Setup:}
We evaluate SEANA on real signal-dependent noise using a real-world fluorescence imaging dataset (BBBC048)\cite{eulenberg2017reconstructing} of Jurkat Cells from the Broad Bioimaging Institute \cite{ljosa2012annotated}. It contains 32{,}266 single-cell flow cytometry images with seven cell-cycle stage labels (\texttt{G1}, \texttt{S}, \texttt{G2}, \texttt{prophase}, 
\texttt{metaphase}, \texttt{anaphase},\texttt{telophase}). This imaging modality exhibits inherent measurement / acquisition noise from photon detection, making this a natural testbed for SEANA. We use a stratified 80/20 train/test split, pretrain for 200 epochs, and evaluate via linear probing on frozen backbones. No noise model or parameters or denoised images are provided to SEANA, it learns the VST entirely from the noisy training images. We compare against baseline SSL methods, and using dataset-specific denoisers (N2N, NSP) with each baseline SSL during pretraining and downstream inference.
\vspace{-10pt}

\paragraph{Results:}
Table~\ref{tab:real_noise} shows SEANA improves all SSL methods. Gains are striking for sample-contrastive and dimensional-contrastive methods. For example, SimCLR jumps from 46.9 to 75.5 (+28.6 pp) and VICReg from 46.8 to 74.4 (+27.5 pp). Notably, all SSL methods converge to similar accuracy under SEANA (74.4-75.7\%). This convergence, despite wide baseline divergence (46.8\%, 63.4\% etc), suggests SEANA eliminates noise as the dominant source of performance variation.
\vspace{-10pt}

\begin{table}[!hb]
\caption{\textbf{Real-world noise linear probing accuracy (\%).} 
For Jurkat cell cycle stage classification, we evaluate SEANA as an augmentation module for joint-embedding SSL. N2N is the Neighbor2Neighbor SSL Denoiser, NSP is the Next-Scale Prediction SSL Denoiser. We report Top-1 accuracy on the frozen backbone across seven cell cycle stage classes (mean $\pm$ std, 3 seeds). $\Delta$ is SEANA's improvement over baseline SSL. Best in \textbf{bold}.}
\label{tab:real_noise}
\centering
\small
\setlength{\tabcolsep}{5pt}
\resizebox{\textwidth}{!}{%
\begin{tabular}{llccccc}
\toprule
\textbf{Noise}
& \textbf{SSL}
& \textbf{Baseline}
& \textbf{N2N (Denoiser) + SSL}
& \textbf{NSP (Denoiser) + SSL}
& \textbf{SEANA (Ours)}
& $\boldsymbol{\Delta}$ \\
\midrule
Real noise
& BYOL
& 63.4\pms{3.4}
& 67.3\pms{2.0}
& 67.9\pms{2.3}
& \textbf{75.7\pms{0.1}}
& \up{12.3} \\

& SimCLR
& 46.9\pms{0.3}
& 48.2\pms{0.6}
& 46.6\pms{0.6}
& \textbf{75.5\pms{0.1}}
& \up{28.6} \\

& VICReg
& 46.8\pms{0.2}
& 47.4\pms{0.8}
& 48.6\pms{0.3}
& \textbf{74.4\pms{0.2}}
& \up{27.5} \\

& LeJEPA
& 47.5\pms{0.4}
& 49.4\pms{0.3}
& 49.3\pms{0.4}
& \textbf{74.4\pms{0.3}}
& \up{26.9} \\

& MoCoV2
& 71.5\pms{1.9}
& 73.3\pms{0.1}
& 73.9\pms{0.1}
& \textbf{75.1\pms{0.6}}
& \up{3.6} \\
\bottomrule
\end{tabular}%
}
\end{table}
\begin{table} 
\caption{\textbf{Ablation study of SEANA components.} Linear probing classification accuracy (\%) on real fluorescence imaging data. We isolate VST and Noise Resampling. Their combination in SEANA provides a significant synergistic boost.}
\label{tab:ablation}
\centering
\small
\begin{tabular}{lcccc}
\toprule
\textbf{SSL Method} & \textbf{Baseline} & \textbf{VST Only} & \textbf{Resampling Only} & \textbf{SEANA (Full)} \\
\midrule
BYOL   & 63.4\pms{3.4} & 59.6\pms{0.1} & 46.5\pms{0.2} & \textbf{75.7\pms{0.1}} \\
SimCLR & 46.9\pms{0.3} & 47.3\pms{0.5} & 74.2\pms{0.0} & \textbf{75.5\pms{0.1}} \\
VICReg & 46.8\pms{0.2} & 46.1\pms{1.0} & 73.3\pms{0.1} & \textbf{74.4\pms{0.2}} \\
LeJEPA & 47.5\pms{0.4} & 47.6\pms{0.3} & 73.2\pms{0.1} & \textbf{74.4\pms{0.3}} \\
MoCoV2 & 71.5\pms{1.9} & 66.8\pms{0.1} & 68.1\pms{1.1} & \textbf{75.1\pms{0.6}} \\
\bottomrule
\end{tabular}
\end{table}

\subsection{Ablation Study: Components of SEANA}
\label{sec:ablation}

We decompose SEANA into its two core components, the learned VST and noise resampling, to isolate their contributions. Using the BBBC048 setup, we evaluate: (1)~\textbf{Baseline}, (2)~\textbf{VST Only} ($\sigma_{\mathrm{vst}}{=}0$), (3)~\textbf{Resampling Only} (identity VST, $\sigma_{\mathrm{vst}}{=}0.8$), and (4)~\textbf{Full SEANA} ($\sigma_{\mathrm{vst}}{=}0.8$). See our analysis of the ablation results in Table \ref{tab:ablation} accompanying Figure \ref{fig:tsne_sweep}.

\FloatBarrier
\noindent\begin{minipage}{\textwidth}
  \begin{minipage}[t]{0.52\textwidth}
    \vspace{0pt} 
    \textbf{Results and Synergies of SEANA componenets:} Table~\ref{tab:ablation} shows that neither component alone suffices. VST-only denoising tends to stay within $\pm$1~pp of baseline for contrastive SSL and degrades non-contrastive SSL, proving that removing noise without reintroducing stochasticity provides minimal benefit. Resampling without the learned VST exhibits a striking asymmetry: it aids explicit and dimensional contrastive methods (SimCLR: 74.2, VICReg: 73.3) but triggers a catastrophic \emph{collapse} in BYOL (46.5\%, $-$16.9~pp drop). As seen in Figure~\ref{fig:tsne_sweep}, BYOL representations degenerate onto a one-dimensional manifold. The model cannot recover from misaligned noise injected in the raw domain. Conversely, Full SEANA improves all frameworks to 75.5--75.7\%. While VST provides modest gains for contrastive models (+1.2~pp), it is critical for BYOL to statistically align resampled noise with original corruptions.
  \end{minipage}
  \hfill
  \begin{minipage}[t]{0.42\textwidth}
    \vspace{0pt}
    \centering
    \includegraphics[width=\linewidth]{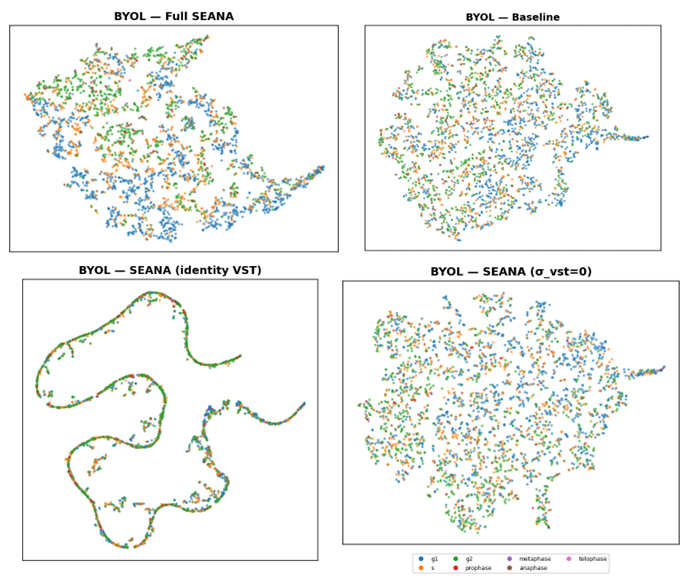}
    \captionsetup{hypcap=false}
    \captionof{figure}{\textbf{BYOL t-SNE ablation:} Identity VST (where resampling only is done) shows manifold collapse compared to full SEANA with more expressive features.}
    \label{fig:tsne_sweep}
  \end{minipage}
\end{minipage}\par

\paragraph{Computational efficiency:}
Table~\ref{tab:comparison} in Appendix ~\ref{app:comp_eff} compares SEANA against the alternative 
of training a dataset-specific denoiser and resampling Gaussian 
noise in the raw domain. SEANA's VST fitting optimizes 130 spline 
parameters over 5{,}000 steps, compared to 1.9M parameters over 
20{,}000 steps for a Neighbor2Neighbor denoiser, a nearly 
15{,}000$\times$ reduction in trainable parameters. Despite this 
efficiency, SEANA is the only approach that generalizes across 
both contrastive and non-contrastive SSL: the train a denoiser and resample
alternative is not only more compute heavy but equivalent to SEANA's ``Resampling Only'' condition, 
which collapses for BYOL.
\vspace{-12pt}

\section{Conclusion, Limitations, and Future Work}
\label{sec:conclusion}

We introduced SEANA, a drop-in module enabling noise-robust representation learning for SSL under signal-dependent noise. By learning a variance-stabilizing transform from noisy data, SEANA reduces the signal-dependent case to a homoscedastic setting. This enables simple Gaussian resampling to approximately satisfy the noise alignment condition of \cite{van2025joint}, without requiring noise model knowledge or dataset-specific denoisers. SEANA assumes pixelwise independent noise, which excludes 
spatially correlated corruptions such as in synthetic aperture radar or structured detector artifacts. The resampling strength $\sigma_{\mathrm{vst}}$ requires tuning per dataset, though we find a simple residual-based heuristic effective in practice. Additionally, SEANA's quality depends on the frozen denoiser's ability to separate signal from noise in the VST domain, such as extremely low-SNR regimes where the denoiser fails may limit gains. Natural extensions include adapting SEANA to spatially correlated noise models, evaluating larger backbones and datasets (e.g., ImageNet-1k with ViT encoders), and exploring integration with reconstruction-based SSL methods such as masked autoencoders.

\clearpage
\begin{ack}
Thank you to Dr. Sriram Vishwanath for aid in providing compuational resources during the rebuttal process while the lead author was transitioning to his PhD program during the summer. Furthermore, thank you to support from the AI Makerspace provided by the Partnership for an Advanced Computing Environment (PACE) at Georgia Tech, Atlanta, Georgia, USA. This research also used the DeltaAI advanced computing and data resource, which is supported by the National Science Foundation (award OAC 2320345) and the State of Illinois. DeltaAI is a joint effort of the University of Illinois Urbana-Champaign and its National Center for Supercomputing Applications.
\end{ack}

\bibliographystyle{plainnat}
\bibliography{example_paper}
\appendix
\clearpage
\section{Monotone CPWL spline parameterization for the VST}
\label{app:cpwl_vst}
We provide further detail on how we parameterize the
piecewise-linear VST $T_\psi$ to guarantee \emph{strict monotonicity} (and hence
invertibility), and how we evaluate $T_\psi$ and $T_\psi^{-1}$ efficiently.

\subsection{Knot placement}
We learn a \emph{scalar}, pointwise intensity transform $T_\psi:[0,1]\rightarrow\mathbb{R}$.
Let $K$ denote the number of knots. We fix knot locations uniformly on the normalized
intensity range:
\begin{equation}
u_k = \frac{k-1}{K-1},\qquad k=1,\dots,K,
\label{eq:cpwl_knots_x}
\end{equation}
and learn knot values $\{y_k\}_{k=1}^K$ that define the spline in VST space.

\subsection{Strictly monotone knot values via exponential increments}
To ensure $T_\psi$ is strictly increasing, it suffices to enforce
\begin{equation}
y_{k+1} - y_k > 0,\qquad \forall k=1,\dots,K-1,
\end{equation}
since $u_{k+1}-u_k>0$ by construction. We achieve this by parameterizing the
\emph{increments} between knot values with an exponential mapping.

Let $\mathbf{s}\in\mathbb{R}^{K-1}$ be unconstrained parameters and define
positive increments
\begin{equation}
\Delta_k \;=\; \exp(s_k) + \varepsilon,\qquad k=1,\dots,K-1,
\label{eq:cpwl_increments}
\end{equation}
where $\varepsilon>0$ is a small constant (e.g., $10^{-6}$) that prevents
numerical issues from extremely small slopes. We then set
\begin{equation}
y_1 = b,\qquad
y_{k+1} = y_k + \Delta_k\;\;\; \text{for}\;\; k=1,\dots,K-1,
\label{eq:cpwl_knot_recursion}
\end{equation}
with a learnable offset $b\in\mathbb{R}$. This guarantees
$y_1<y_2<\cdots<y_K$, so $T_\psi$ is strictly increasing and therefore invertible.

\paragraph{Optional range normalization:}
If desired, one can constrain the VST output to a finite interval $[y_{\min},y_{\max}]$
by re-scaling the cumulative sums:
\begin{equation}
y_k \;=\; y_{\min} \;+\; \frac{\sum_{j=1}^{k-1}\Delta_j}{\sum_{j=1}^{K-1}\Delta_j}\,(y_{\max}-y_{\min}),
\qquad k=1,\dots,K,
\label{eq:cpwl_range_norm}
\end{equation}
(with the convention $\sum_{j=1}^0 \Delta_j = 0$). In our implementation we allow
$y_k$ to be unconstrained (Eq.~\ref{eq:cpwl_knot_recursion}), and handle values outside
the knot range by linear extrapolation (below).

\subsection{Forward evaluation of \texorpdfstring{$T_\psi$}{T(psi)}}
For any input $x\in[0,1]$, let $k$ be the (unique) index such that
$x\in[u_k,u_{k+1}]$. The CPWL spline is defined by linear interpolation:
\begin{equation}
T_\psi(x) \;=\; y_k \;+\; \frac{x-u_k}{u_{k+1}-u_k}\,(y_{k+1}-y_k),
\qquad x\in[u_k,u_{k+1}].
\label{eq:cpwl_forward}
\end{equation}
We apply this pointwise to every pixel/channel.

\paragraph{Tails (out-of-range handling):}
During augmentation, VST-space noise injection can produce values outside
$[y_1,y_K]$ after inversion steps; likewise, numerical normalization can produce
$x$ slightly outside $[0,1]$. We therefore use \emph{linear tails}:
\begin{align}
T_\psi(x) &=
\begin{cases}
y_1 + (x-u_1)\,m_1, & x<u_1,\\
\text{Eq.~\ref{eq:cpwl_forward}}, & x\in[u_1,u_K],\\
y_K + (x-u_K)\,m_{K-1}, & x>u_K,
\end{cases}
\label{eq:cpwl_forward_tails}
\end{align}
where $m_k=\frac{y_{k+1}-y_k}{u_{k+1}-u_k}$ is the slope of segment $k$.

\subsection{Inverse evaluation of \texorpdfstring{$T_\psi^{-1}$}{T(psi) inverse}}
Because $T_\psi$ is strictly increasing, its inverse exists and is also piecewise-linear.
Given $t\in\mathbb{R}$, find the index $k$ such that $t\in[y_k,y_{k+1}]$ and compute
\begin{equation}
T_\psi^{-1}(t) \;=\; u_k \;+\; \frac{t-y_k}{y_{k+1}-y_k}\,(u_{k+1}-u_k),
\qquad t\in[y_k,y_{k+1}].
\label{eq:cpwl_inverse}
\end{equation}
We again use linear tails for $t<y_1$ and $t>y_K$:
\begin{align}
T_\psi^{-1}(t) &=
\begin{cases}
u_1 + (t-y_1)/m_1, & t<y_1,\\
\text{Eq.~\ref{eq:cpwl_inverse}}, & t\in[y_1,y_K],\\
u_K + (t-y_K)/m_{K-1}, & t>y_K.
\end{cases}
\label{eq:cpwl_inverse_tails}
\end{align}

\subsection{Affine correction in the inverse map}
We optionally absorb small systematic biases induced by
approximate stabilization by applying a lightweight affine correction after inversion:
\begin{equation}
T_\psi^{-1}(t)\;\leftarrow\; T_\psi^{-1}(t) + \alpha t + \beta,
\end{equation}
with learnable scalars $(\alpha,\beta)$.

\section{Synthetic Noise Models}
\label{app:noise_models}

For synthetic experiments, we corrupt clean images with three signal-dependent noise models, each applied independently per pixel and channel with the result clipped to $[0,1]$. In all cases, $\boldsymbol{\epsilon} \sim 
\mathcal{N}(\mathbf{0}, \mathbf{I})$.

\textbf{Additive Gaussian Noise:}
$\tilde{\mathbf{x}} = \operatorname{clip}(
\mathbf{x} + \sigma\,\boldsymbol{\epsilon},\; 0,\; 1)$,
where $\sigma > 0$ controls the noise level.

\textbf{Poisson--Gaussian (Shot/Read) Noise:} Models 
signal-dependent shot noise via a Poisson process with additive read noise:
$\tilde{\mathbf{x}} = \operatorname{clip}(
\operatorname{Poisson}(\mathbf{x} \cdot \lambda)/\lambda 
+ \sigma_g\,\boldsymbol{\epsilon},\; 0,\; 1)$,
where $\lambda > 0$ controls the photon count and 
$\sigma_g \geq 0$ the read noise.

\textbf{Multiplicative (Speckle) Noise:}
$\tilde{\mathbf{x}} = \operatorname{clip}(
\mathbf{x} \odot (1 + \sigma\,\boldsymbol{\epsilon}),\; 0,\; 1)$,
where $\sigma > 0$ controls the speckle strength.

\section{Real Clinical Imaging Experiment: SSL of Fetal Ultrasound Data}
\label{app:fetal_exp}

\paragraph{Dataset and experimental setup:}
To evaluate SEANA beyond fluorescence imaging, we use the Fetal
Planes DB dataset~\cite{burgos2020evaluation}. Ultrasound forms images using
sound waves, providing a different acquisition setting from BBBC048.
The dataset contains 12,400 images from 1,792 patients, collected across
different hospitals, operators, and ultrasound machines. Images belong
to six classes: fetal abdomen, brain, femur, thorax, maternal cervix,
and other planes.

We follow the official patient-disjoint split, with 7,129 training
images from 896 patients and 5,271 test images from 896 different
patients. We compare baseline SSL, NSP denoising followed by SSL,
and SEANA across SimCLR, VICReg, LeJEPA, BYOL, and MoCoV2 \cite{balestriero2025lejepa} \cite{chen2020improved}. For SEANA, we used DRUNet with $\sigma_d$ = 0.08 across all SSL methods and $\sigma_\text{vst}$ = 0.16 (SimCLR), 0.16 (VICReg), 0.24 (LeJEPA), 0.12 (BYOL), 0.04 (MoCoV2). No synthetic
noise is added to the dataset itself. After SSL pretraining, we evaluate each frozen backbone
using a linear classifier. Our primary metric is macro-F1, which gives
equal weight to each class, averaged over three training seeds.

\paragraph{Overall performance:}
SEANA achieves the highest mean macro-F1 for all five SSL methods
(Table~\ref{tab:fetal_macro_f1}). The largest gains occur for SimCLR,
VICReg, and LeJEPA. BYOL and MoCoV2 start from stronger baselines and
show smaller changes, with MoCoV2 remaining approximately unchanged.
NSP produces mixed results, indicating that denoising alone does not
consistently improve the learned representations.

\begin{table}[t]
    \centering
    \small
    \caption{Fetal Planes DB linear-probe macro-F1 (\%).
    Results are means $\pm$ sample standard deviations across three
    seeds. Standard deviations are in percentage points.}
    \label{tab:fetal_macro_f1}
    \begin{tabular}{lccc}
        \toprule
        SSL method & Baseline & NSP + SSL & SEANA \\
        \midrule
        SimCLR & $65.57\pm1.20$ & $66.64\pm3.87$ & $71.01\pm1.43$ \\
        VICReg & $65.39\pm1.88$ & $64.55\pm2.65$ & $68.79\pm0.20$ \\
        LeJEPA & $67.41\pm0.32$ & $68.15\pm0.82$ & $69.22\pm0.41$ \\
        BYOL   & $71.11\pm0.67$ & $71.44\pm1.19$ & $72.06\pm0.82$ \\
        MoCoV2 & $69.11\pm0.77$ & $68.77\pm0.12$ & $69.24\pm0.73$ \\
        \bottomrule
    \end{tabular}
\end{table}

\paragraph{Anatomical breakdown: }
The per-class results in Table \ref{tab:fetal_anatomy} show substantial improvements for some of the
classes with lower baseline performance. For SimCLR, SEANA increases
abdomen F1 from $44.0\%$ to $55.7\%$ and femur F1 from $43.0\%$ to
$51.8\%$. VICReg similarly improves abdomen F1 from $42.1\%$ to
$51.4\%$. Improvements are not uniform across anatomical classes,
highlighting the value of examining per-class performance alongside
macro-F1.

\begin{table*}[t]
\caption{
\textbf{Per-class anatomical classification on Fetal Planes DB:}
We report per-class F1 and macro-F1 (\%) from frozen-backbone linear
probes, averaged across three seeds.
Cervix refers to the maternal cervix.
For each SSL method, the higher value between Baseline and SEANA
in each column is shown in \textbf{bold}.
}
\label{tab:fetal_anatomy}
\centering
\small
\setlength{\tabcolsep}{4pt}
\resizebox{\textwidth}{!}{%
\begin{tabular}{llrrrrrrr}
\toprule
\textbf{SSL Method}
& \textbf{Approach}
& \textbf{Brain}
& \textbf{Thorax}
& \textbf{Cervix}
& \textbf{Abdomen}
& \textbf{Femur}
& \textbf{Other}
& \textbf{Macro-F1} \\
\midrule

SimCLR
& Baseline
& 78.7 & 65.5 & \textbf{92.5} & 44.0 & 43.0 & 69.7 & 65.6 \\
& SEANA
& \textbf{84.3} & \textbf{70.5} & 91.5 & \textbf{55.7}
& \textbf{51.8} & \textbf{72.2} & \textbf{71.0} \\
\midrule

VICReg
& Baseline
& 77.3 & 66.5 & \textbf{92.1} & 42.1 & 44.2 & 70.2 & 65.4 \\
& SEANA
& \textbf{82.3} & \textbf{70.1} & 90.4 & \textbf{51.4}
& \textbf{47.8} & \textbf{70.8} & \textbf{68.8} \\
\midrule

LeJEPA
& Baseline
& 78.3 & 69.5 & \textbf{91.6} & 43.3
& \textbf{51.2} & 70.6 & 67.4 \\
& SEANA
& \textbf{82.8} & \textbf{70.2} & 90.7 & \textbf{51.8}
& 48.9 & \textbf{70.9} & \textbf{69.2} \\
\midrule

MoCoV2
& Baseline
& 81.6 & 70.0 & \textbf{91.8} & \textbf{50.1}
& 50.4 & 70.8 & 69.1 \\
& SEANA
& \textbf{81.9} & \textbf{70.5} & 91.0 & 48.6
& \textbf{52.3} & \textbf{71.1} & \textbf{69.2} \\
\midrule

BYOL
& Baseline
& 82.1 & 71.8 & 95.9 & 51.2 & \textbf{53.6} & 72.1 & 71.1 \\
& SEANA
& \textbf{83.7} & \textbf{72.8} & \textbf{96.0} & \textbf{55.2}
& 51.8 & \textbf{72.9} & \textbf{72.1} \\

\bottomrule
\end{tabular}%
}
\end{table*}

\paragraph{Performance across ultrasound machines: }
The patient split also contains a substantial device imbalance.
Voluson S10 contributes only 11 training images ($0.15\%$ of training)
but 1,071 test images ($20.3\%$ of testing), making it a nearly unseen
acquisition setting.

In device acquisition distribution shift analysis in Table \ref{tab:fetal_devices}, SEANA improves S10 accuracy for
all five SSL methods, with larger gains on the uncommon S10 against the existing commonly observed
E6 and Aloka devices (Table~\ref{tab:fetal_devices}). For example,
SimCLR improves from $52.26\%$ to $59.23\%$ on S10, while LeJEPA
improves from $51.17\%$ to $60.88\%$.

\begin{table}[t]
    \centering
    \small
    \caption{Acquisition distribution shift by device-stratified accuracy (\%). E6 and Aloka are in-distribution ultrasound machines Volusion S10 is an acquisition distribution shift ultrasound machine not reflected in the training data.
    Each entry shows baseline $\rightarrow$ SEANA.}
    \label{tab:fetal_devices}
    \begin{tabular}{lcc}
        \toprule
        SSL method & E6 + Aloka & Voluson S10 \\
        \midrule
        SimCLR & $76.44\rightarrow \textbf{79.50}$ & $52.26\rightarrow\textbf{59.23}$ \\
        VICReg & $76.01\rightarrow\textbf{77.50}$ & $52.10\rightarrow\textbf{57.98}$ \\
        LeJEPA & $\textbf{77.58}\rightarrow 76.90$ & $51.17\rightarrow\textbf{60.88}$ \\
        BYOL   & $79.77\rightarrow\textbf{80.11}$ & $57.21\rightarrow\textbf{62.00}$ \\
        MoCoV2 & $77.01\rightarrow\textbf{77.21}$ & $60.85\rightarrow\textbf{61.78}$ \\
        \bottomrule
    \end{tabular}
\end{table}

Together with the BBBC048 experiments, these results demonstrate SEANA's applicability across fluorescence and ultrasound imaging. The magnitude of improvement varies across SSL methods, with the device analysis suggesting particular benefits under changes in acquisition conditions.

\section{Ablation of SEANA with Frozen, Off-The-Shelf Gaussian Denoisers}
\label{app:den_abl}

Since SEANA leverages a blind-spot frozen, off-the-shelf denoiser as part of its clean-target estimate, to check if our method generalizes, we ablate SEANA with a number of different frozen, off-the-shelf denoisers on BBBC048 (Real-World Fluoresence Microscopy Data). In the main paper, we used DRUNet.

\begin{table}[b]
\caption{\textbf{Denoiser generalization:} SEANA with four 
different frozen, off-the-shelf blind-spot denoisers on BBBC048. All achieve similar 
accuracy, confirming SEANA is architecture-agnostic. 
Mean $\pm$ std over 3 seeds.}
\label{tab:denoiser_ablation}
\centering
\small
\setlength{\tabcolsep}{5pt}
\begin{tabular}{llcccc}
\toprule
& & \textbf{DRUNet} & \textbf{DnCNN} & \textbf{SCUNet} & \textbf{SwinIR} \\
\textbf{SSL} & \textbf{Base} & 
\scriptsize U-Net+Res & \scriptsize Plain CNN & 
\scriptsize Swin-UNet & \scriptsize Swin Trans. \\
\midrule
BYOL   & 63.4\pms{3.4} & 75.7\pms{0.1} & 75.3\pms{0.1} & \textbf{75.8\pms{0.1}} & 75.3\pms{0.1} \\
SimCLR & 46.9\pms{0.3} & \textbf{75.5\pms{0.1}} & 75.1\pms{0.1} & 75.3\pms{0.1} & 75.2\pms{0.1} \\
VICReg & 46.8\pms{0.2} & 74.4\pms{0.2} & 74.7\pms{0.1} & \textbf{75.2\pms{0.2}} & 75.1\pms{0.2} \\
\midrule
\textit{Mean} & \textit{52.4} & \textit{75.2} & \textit{75.0} & \textit{\textbf{75.4}} & \textit{75.2} \\
\bottomrule
\end{tabular}
\end{table}

Overall, across all gaussian denoiser families (DRUNet, DnCNN, SCUNet, and SwinIR), all achieve similar accuracy (75.0--75.4\% mean across SSL methods), confirming that SEANA's gains arise from noise-aligned resampling via the learned VST, not from the denoising quality of any particular architecture.

\section{Comparison with known-noise and oracle augmentation}
\label{app:oracle_comparison}

\paragraph{Experimental setup:}
We examine how SEANA compares with augmentation methods that have access
to the true noise model or clean images. We evaluate BYOL, SimCLR, and
VICReg under the same six synthetic CIFAR-10 noise conditions used in
the main experiments: Gaussian, Poisson--Gaussian, and multiplicative
noise, each at low and high strength. We measure clean-test linear-probe
accuracy, with the classifier trained on noisy images. Summary results
average equally over the 18 SSL-method/noise-condition combinations.

We consider the following controls:
\begin{itemize}
    \item \textbf{Oracle noise-aligned augmentation:}
    We generate independently corrupted views from the true clean image
    using the exact noise model. This provides a privileged reference
    with access to both clean content and the corruption process.

    \item \textbf{Analytic VST plus resampling:}
    We replace SEANA's learned transform with a fixed transform chosen
    for the noise family: identity for Gaussian noise, a generalized
    Anscombe transform for Poisson--Gaussian noise, and a logarithmic
    transform for multiplicative noise.

    \item \textbf{Denoising plus raw-domain resampling:}
    We first estimate the clean image using NSP or N2N, then independently
    corrupt each view using the true noise model in the original image
    domain. We also include NSP and N2N without resampling.
\end{itemize}

\begin{table*}[t]
    \centering
    \small
    \setlength{\tabcolsep}{4pt}
    \caption{
        Summary across the 18 SSL-method/noise-condition combinations on synthetic noisy CIFAR-10.
        Accuracy is reported in percent and gains in percentage points
        (pp) relative to the corresponding baseline.
        R denotes independent noise resampling.
        Privileged information refers to access to clean training images
        or knowledge of the true corruption process.
    }
    \label{tab:oracle_summary}
    \resizebox{\textwidth}{!}{%
    \begin{tabular}{cllrrrr}
        \toprule
        Rank
        & Approach
        & \shortstack[l]{Privileged\\information}
        & \shortstack{Mean\\accuracy (\%)}
        & \shortstack{Mean gain\\(pp)}
        & \shortstack{Median gain\\(pp)}
        & \shortstack{Gain range\\(pp)} \\
        \midrule
        --
        & Oracle
        & \shortstack[l]{Clean images and\\exact noise model}
        & 59.37 & $+16.69$ & $+16.50$ & $[+7.0,+27.0]$ \\

        1
        & \textbf{SEANA}
        & \textbf{None}
        & 57.68 & $+15.01$ & $+14.15$ & $[+5.9,+26.6]$ \\

        2
        & Analytic VST + R
        & \shortstack[l]{Noise family and\\relevant parameters}
        & 54.86 & $+12.18$ & $+12.85$ & $[+1.8,+22.8]$ \\

        3
        & NSP + R
        & Exact noise model
        & 53.24 & $+10.56$ & $+11.55$ & $[-2.0,+19.8]$ \\

        4
        & NSP
        & None
        & 46.35 & $+3.67$ & $+2.25$ & $[-1.9,+16.4]$ \\

        5
        & N2N + R
        & Exact noise model
        & 45.68 & $+3.00$ & $+4.65$ & $[-8.6,+14.0]$ \\

        6
        & N2N
        & None
        & 43.49 & $+0.82$ & $+0.20$ & $[-4.6,+7.0]$ \\

        --
        & Baseline
        & None
        & 42.68 & $0.00$ & $0.00$ & $[0.0,0.0]$ \\
        \bottomrule
    \end{tabular}%
    }
\end{table*}

\paragraph{Results:}
SEANA achieves the highest mean accuracy among methods without access
to clean training images: $57.68\%$, compared with $54.86\%$ for analytic
VST plus resampling, $53.24\%$ for NSP plus resampling, and $45.68\%$
for N2N plus resampling (Table~\ref{tab:oracle_summary}). Its mean gain
over baseline is $15.01$ percentage points, approximately $90\%$ of
the oracle's mean gain. SEANA remains about $1.7$ percentage points
below the oracle's $59.37\%$ mean accuracy.

The full results show that SEANA outperforms both denoiser-plus-resampling
baselines in all 18 comparisons and the analytic VST control in 16 of
18 comparisons (Table~\ref{tab:oracle_full}). SEANA improves over baseline
in every setting, with gains ranging from $5.9$ to $26.6$ percentage
points. The remaining gap to the oracle is largest under high Gaussian
noise, averaging $5.7$ percentage points across the three SSL methods.

\begin{table*}[t]
    \centering
    \small
    \setlength{\tabcolsep}{3pt}
    \caption{
        Clean-test linear-probe accuracy (\%) for each synthetic CIFAR-10
        noise condition and SSL method. Probes are trained on noisy images.
        R denotes independent noise resampling.
        NSP + R and N2N + R resample using the exact noise model in the
        original image domain. Analytic VST + R resamples in the
        transformed domain. PG denotes Poisson--Gaussian noise.
    }
    \label{tab:oracle_full}
    \resizebox{\textwidth}{!}{%
    \begin{tabular}{llrrrrrrrr}
        \toprule
        Noise condition
        & SSL
        & N2N
        & NSP
        & N2N + R
        & NSP + R
        & \shortstack{Analytic\\VST + R}
        & SEANA
        & Oracle
        & Baseline \\
        \midrule

        Gaussian (low)
        & BYOL
        & 51.8 & 54.4 & 57.6 & 57.4 & 60.7 & 60.8 & 61.4 & 52.3 \\
        & SimCLR
        & 59.8 & 58.6 & 62.3 & 58.1 & 61.9 & 66.0 & 67.2 & 60.1 \\
        & VICReg
        & 64.8 & 61.7 & 67.3 & 64.6 & 69.3 & 69.4 & 69.5 & 62.5 \\
        \midrule

        Gaussian (high)
        & BYOL
        & 22.9 & 25.8 & 26.9 & 40.3 & 41.0 & 41.9 & 49.3 & 27.5 \\
        & SimCLR
        & 28.1 & 29.2 & 36.1 & 40.5 & 42.5 & 42.2 & 49.1 & 22.1 \\
        & VICReg
        & 30.1 & 28.9 & 38.3 & 43.8 & 43.4 & 47.4 & 50.1 & 30.8 \\
        \midrule

        PG (low)
        & BYOL
        & 48.0 & 48.9 & 52.9 & 56.8 & 59.6 & 59.9 & 59.9 & 46.5 \\
        & SimCLR
        & 53.5 & 57.4 & 58.7 & 56.9 & 61.8 & 65.0 & 64.8 & 51.1 \\
        & VICReg
        & 56.9 & 60.0 & 64.6 & 64.3 & 68.7 & 69.1 & 69.7 & 58.7 \\
        \midrule

        PG (high)
        & BYOL
        & 27.5 & 35.7 & 23.8 & 45.7 & 44.5 & 47.8 & 51.6 & 29.7 \\
        & SimCLR
        & 29.2 & 37.6 & 28.6 & 47.0 & 41.1 & 50.5 & 52.3 & 27.2 \\
        & VICReg
        & 31.1 & 33.0 & 30.3 & 51.8 & 53.0 & 54.7 & 54.7 & 32.1 \\
        \midrule

        Speckle (low)
        & BYOL
        & 51.3 & 51.2 & 56.6 & 57.7 & 59.0 & 61.5 & 61.6 & 52.1 \\
        & SimCLR
        & 58.8 & 59.8 & 63.1 & 58.8 & 59.5 & 67.0 & 67.4 & 55.5 \\
        & VICReg
        & 61.0 & 63.2 & 68.8 & 66.6 & 67.3 & 71.4 & 71.7 & 63.7 \\
        \midrule

        Speckle (high)
        & BYOL
        & 34.4 & 36.1 & 29.2 & 46.8 & 46.8 & 50.1 & 54.0 & 30.3 \\
        & SimCLR
        & 35.1 & 44.5 & 27.8 & 47.9 & 46.7 & 54.7 & 54.5 & 28.1 \\
        & VICReg
        & 38.6 & 48.3 & 29.3 & 53.3 & 60.7 & 58.9 & 59.8 & 37.9 \\
        \bottomrule
    \end{tabular}%
    }
\end{table*}

\begin{table}[t]
\caption{
\textbf{Denoising quality on CIFAR-10 under high synthetic noise.}
PSNR is reported in dB ($\uparrow$), computed from pooled pixel-wise
mean squared error. DRUNet + SEANA VST evaluates the estimated
clean image before noise resampling.
Best results are shown in \textbf{bold}.
}
\label{tab:denoiser_psnr}
\centering
\small
\setlength{\tabcolsep}{5pt}
\begin{tabular}{lcccc}
\toprule
\textbf{Noise condition}
& \textbf{N2N}
& \textbf{NSP}
& \textbf{DRUNet}
& \shortstack{\textbf{DRUNet + \textbf{SEANA} VST}} \\
\midrule
Gaussian (high)
& 13.46 & 17.55 & 16.90 & \textbf{17.96} \\

Poisson--Gaussian (high)
& 15.00 & 18.44 & 15.97 & \textbf{19.96} \\

Speckle (high)
& 16.00 & 18.76 & 13.56 & \textbf{20.13} \\
\bottomrule
\end{tabular}
\end{table}

These results show that knowing the corruption process alone does not
guarantee better SSL representations. Under the evaluated settings,
SEANA's learned transform and resampling provide stronger average
performance than either a fixed analytic transform or denoising
followed by exact noise resampling, without requiring clean training
images or access to the true noise model.

\section{Experiments Implementation Details}
\label{app:impl_details}
\paragraph{Dataset Details:} We leverage the ImageNet-100 and CIFAR-10 datasets and add synthetic noise to them, following the synthetic noise models described in Appendix \ref{app:noise_models}. This is a deterministic process done once offline, producing 6 different synthetically noisy datasets for each clean dataset we started with. The BBBC048 dataset \cite{eulenberg2017reconstructing} is publicly available from the Broad Bioimage Benchmark Collection \cite{ljosa2012annotated}. It contains 32{,}266 single-cell fluorescence microscopy images of Jurkat cells captured during cell cycle progression. Each image is labeled with one of seven cell cycle phases: \texttt{G1}, 
\texttt{S}, \texttt{G2}, \texttt{prophase} \texttt{metaphase}, \texttt{anaphase}, and \texttt{telophase}. For BBBC048, we use pre-merged 3-channel composite images (combining fluorescence channels Ch3/Ch4/Ch6), which are provided in the dataset under \texttt{CellCycle/<Phase>/} directories. Images are resized to $64 \times 64$ pixels using bilinear interpolation with center cropping, and normalized to 
$[0,1]$. We apply a stratified 80/20 train/test split 
(25{,}813 training / 6{,}453 test images), stratified by 
phase label with a fixed random seed for reproducibility.

\paragraph{Training Details:} We train each SSL model baseline and optional SEANA integration or N2N Denoiser on an H100 GPU. Each SSL model is trained for 100 epochs with a batch size of 256 on the noisy training data for ImageNet-100, and 200 epochs for CIFAR-10 and BBBC048 on their respective train splits. A linear probe is trained on the frozen SSL backbone for 100 epochs. We specifically look at both contrastive and non-contrastive (BYOL, SimCLR, VICReg) approaches \cite{chen2020simple}, \cite{grill2020bootstrap}, \cite{bardes2021vicreg}.

\paragraph{SSL Model \& Augmentation Details:}
For the data augmentation pipelines and SSL modules, we use default modules from the lightlySSL library. For SimCLR, we use a ResNet-50 backbone and we set the temperature parameter to $\tau$ = 0.5. For VICReg, we use the following default hyperparameters: a scaling coefficient of 25 for the invariance term of the loss, 25 for the variance term, and 1 for the covariance term. For BYOL, we use a ResNet-50 backbone and we use a cosine scheduler from 0.99 to 1 for the student momentum parameter. For augmentations, we use the default augmentations associated to SimCLR, BYOL and VICReg from the lightly library. For CIFAR-10 specifically, we adjust the random resized crop to be set to 32.

\paragraph{Gaussian Denoiser \& SEANA Details:} For the concrete Gaussian Denoiser we use across all experiments, we leverage DruNet \cite{zhang2021plug}. We use a $\sigma_g = 0.09803921568$ ($\sigma=$25/255) across experiments 1-2, and a $\sigma=0.039$ for experiment 3. For gaussian noise injection in SEANA's VST space, we estimate $\sigma_{\text{vst}}$ from the empirical residual standard deviation in VST space, $\widehat{\mathbf{t}}- \mathbf{t}$ and then did sweeps of scaled factors of that estimate, landing on  $\sigma{\mathrm{vst}} = 0.8$ (see Appendix~\ref{app:sigmasense} for sensitivity 
analysis). SEANA is fit on each dataset for 5000 steps and uses a patch size of 64 for CIFAR-10 and BBBC048 and a patch size of 224 for ImageNet-100.

\section{Sensitivity to Spline Knot Count}
\label{app:knot_sweep}

Table~\ref{tab:knot_sweep} reports linear probing accuracy on 
BBBC048 as the number of spline knots $K$ varies from 8 to 128 
(corresponding to 10--130 total trainable parameters). Performance 
is remarkably stable: even $K = 8$ (10 parameters) achieves 
within 1.4~pp of $K = 128$ across all SSL methods. This confirms 
that the variance-stabilizing transform is a simple monotonic 
function that does not require high-resolution parameterization. 
In practice, the default $K = 128$ is a safe choice, but 
practitioners operating under extreme parameter budgets can 
reduce to $K = 16$ with negligible performance loss.

\begin{table}[t]
\caption{\textbf{Knot count sensitivity on BBBC048:} Linear 
probing accuracy (\%) as a function of the number of spline 
knots $K$. Total trainable parameters = $K + 2$ (knot values 
plus affine correction). Seed 0.}
\label{tab:knot_sweep}
\centering
\small
\begin{tabular}{rccccc}
\toprule
& \multicolumn{5}{c}{\textbf{Number of knots} $K$} \\
\cmidrule(l){2-6}
\textbf{SSL} & \textbf{8} & \textbf{16} & \textbf{32} & 
\textbf{64} & \textbf{128} \\
\midrule
BYOL   & 74.4 & \textbf{75.8} & 75.7 & 75.5 & 75.5 \\
SimCLR & 75.4 & 75.5 & \textbf{75.8} & 75.7 & 75.7 \\
VICReg & 74.0 & 74.1 & 73.7 & 74.3 & \textbf{74.4} \\
\midrule
\textit{Mean} & \textit{74.6} & \textit{\textbf{75.1}} & 
\textit{75.1} & \textit{75.2} & \textit{75.2} \\
\bottomrule
\end{tabular}
\end{table}

\section{Computational Efficiency: SEANA vs Dataset-Specific Denoiser with Resampling}
\label{app:comp_eff}

While an off-the-shelf gaussian denoiser with resampling requires no training at all, it provides no guarantee of adaption to the unknown noise distribution and fails for non-contrastive SSL like BYOL. Thus, we qualitatively and quantiatively compare training a dataset-specific denoiser (N2N) with resampling against SEANA in Table \ref{tab:comparison}. 
\begin{table}[t]
\caption{\textbf{Qualitative and quantitative comparison of training pipelines.} SEANA is significantly more parameter-efficient than N2N while maintaining universal framework compatibility.}
\label{tab:comparison}
\centering
\small
\setlength{\tabcolsep}{4pt} 
\begin{tabularx}{\textwidth}{l l X} 
\toprule
& \textbf{SEANA} & \textbf{Dataset-Specific Denoiser (N2N) + Resampling} \\
\midrule
What is optimized              & Monotonic spline (VST) & U-Net (denoiser) \\
Trainable parameters           & \textbf{130}           & 1,259,475 ($\approx 9,700\times$) \\
Optimization steps             & \textbf{5,000}         & 20,000 ($4\times$) \\
Requires clean/paired data     & No                     & No \\
Requires noise model knowledge & No                     & Yes \\
\bottomrule
\end{tabularx}
\vspace{-12pt}
\end{table}

\section{Hyperparameter Sensitivity Analysis of SEANA's \texorpdfstring{$\sigma_{\mathrm{vst}}$}{sigma-VST}}
\label{app:sigmasense}

\begin{figure}[t]
\centering
\includegraphics[width=0.75\textwidth]{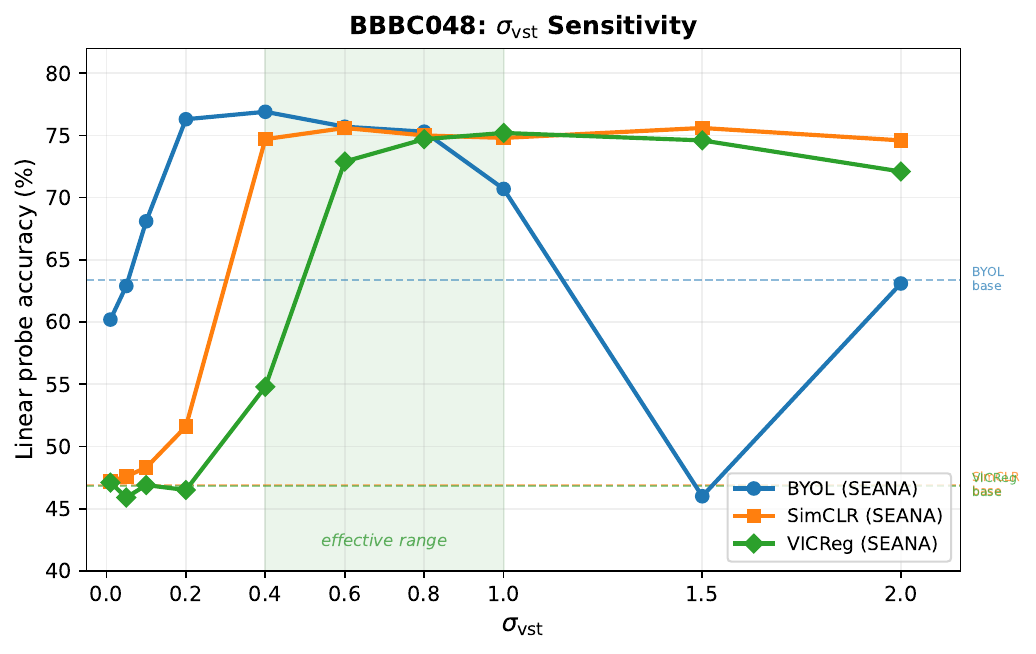}
\caption{\textbf{$\sigma_{\mathrm{vst}}$ sensitivity on BBBC048.} 
Linear probing accuracy vs.\ resampling strength for BYOL, 
SimCLR, and VICReg. Dashed lines indicate baselines without 
SEANA. The shaded region marks the effective range where all 
three methods improve substantially. BYOL is the most sensitive, 
collapsing at $\sigma_{\mathrm{vst}} = 1.5$, while contrastive 
methods maintain a broad plateau.}
\label{fig:sigma_sensitivity}
\end{figure}

Figure~\ref{fig:sigma_sensitivity} reports linear probing 
accuracy on BBBC048 as a function of SEANA's resampling strength 
$\sigma_{\mathrm{vst}}$ for all three SSL methods (seed 0). 
All methods improve substantially over their respective 
baselines (dashed lines) for 
$\sigma_{\mathrm{vst}} \in [0.4, 1.0]$, and the three curves 
converge to $\sim$75\% within this range. Below 
$\sigma_{\mathrm{vst}} = 0.2$, accuracy remains near baseline, 
confirming that sufficient resampling noise is necessary to 
break the noise shortcut.

BYOL exhibits the narrowest effective range, peaking at 
$\sigma_{\mathrm{vst}} = 0.4$ (76.9\%) and collapsing at 
$\sigma_{\mathrm{vst}} = 1.5$ (46.0\%). Without negative pairs 
to regularize learning, excessive resampled noise destroys the 
BYOL training signal. SimCLR and VICReg are more robust, 
maintaining $>$72\% accuracy from $\sigma_{\mathrm{vst}} = 0.4$ 
through 2.0.

\newpage
\section*{NeurIPS Paper Checklist}

\begin{enumerate}

\item {\bf Claims}
    \item[] Question: Do the main claims made in the abstract and introduction accurately reflect the paper's contributions and scope?
    \item[] Answer: \answerYes{} 
    \item[] Justification: The abstract and introduction clearly state the claims, specifically the theoretical motivation for noise-aligned view generation, the proposed SEANA method, and the empirical improvements across three SSL methods and various noise families. The scope of these claims is supported by the experiments detailed in Section~\ref{sec:experiments}.

\item {\bf Limitations}
    \item[] Question: Does the paper discuss the limitations of the work performed by the authors?
    \item[] Answer: \answerYes{} 
    \item[] Justification: Limitations are explicitly discussed in Section~\ref{sec:conclusion}, including the assumption of pixelwise independent noise (which excludes spatially correlated corruptions), the need to tune the resampling strength hyperparameter, and performance dependence on the frozen denoiser in extremely low-SNR regimes.

\item {\bf Theory assumptions and proofs}
    \item[] Question: For each theoretical result, does the paper provide the full set of assumptions and a complete (and correct) proof?
    \item[] Answer: \answerNA{} 
    \item[] Justification: The paper leverages existing theoretical frameworks to motivate the method but does not introduce or prove new formal theorems of its own.
    \item[] Guidelines:

    \item {\bf Experimental result reproducibility}
    \item[] Question: Does the paper fully disclose all the information needed to reproduce the main experimental results of the paper to the extent that it affects the main claims and/or conclusions of the paper (regardless of whether the code and data are provided or not)?
    \item[] Answer: \answerYes{} 
    \item[] Justification: We provide detailed implementation specifics, including dataset generation, noise models, training hyperparameters, and model architectures in Appendix~\ref{app:impl_details} and Appendix~\ref{app:noise_models}.

\item {\bf Open access to data and code}
    \item[] Question: Does the paper provide open access to the data and code, with sufficient instructions to faithfully reproduce the main experimental results, as described in supplemental material?
    \item[] Answer: \answerNo{} 
    \item[] Justification: Code is not provided during the initial double-blind review phase. However, as stated in the abstract, we are committed to making the full code, including the VST calibration and SSL view generation modules, publicly available upon acceptance.

\item {\bf Experimental setting/details}
    \item[] Question: Does the paper specify all the training and test details (e.g., data splits, hyperparameters, how they were chosen, type of optimizer) necessary to understand the results?
    \item[] Answer: \answerYes{} 
    \item[] Justification: Comprehensive experimental settings, including data splits, noise injection procedures, SSL model hyperparameters (like temperature and scaling coefficients), batch sizes, and epochs, are detailed in Appendix~\ref{app:impl_details}.

\item {\bf Experiment statistical significance}
    \item[] Question: Does the paper report error bars suitably and correctly defined or other appropriate information about the statistical significance of the experiments?
    \item[] Answer: \answerYes{} 
    \item[] Justification: We report the mean and standard deviation over 3 random seeds for all main linear probing accuracy results (Tables~\ref{tab:cifar10_split}, \ref{tab:imagenet100_split}, \ref{tab:real_noise}, and \ref{tab:ablation}).
    \item[] Guidelines:

\item {\bf Experiments compute resources}
    \item[] Question: For each experiment, does the paper provide sufficient information on the computer resources (type of compute workers, memory, time of execution) needed to reproduce the experiments?
    \item[] Answer: \answerYes{} 
    \item[] Justification: The compute hardware (H100 GPU) is specified in Appendix~\ref{app:impl_details}. We also discuss the computational efficiency and optimization steps compared to baselines in Section~\ref{sec:ablation} and Appendix~\ref{app:comp_eff}.
    \item[] Guidelines:
    
\item {\bf Code of ethics}
    \item[] Question: Does the research conducted in the paper conform, in every respect, with the NeurIPS Code of Ethics \url{https://neurips.cc/public/EthicsGuidelines}?
    \item[] Answer: \answerYes{} 
    \item[] Justification: The research conforms to the NeurIPS Code of Ethics. We use standard, publicly available scientific and natural imaging datasets.

\item {\bf Broader impacts}
    \item[] Question: Does the paper discuss both potential positive societal impacts and negative societal impacts of the work performed?
    \item[] Answer: \answerNo{} 
    \item[] Justification: Our work focuses on foundational representation learning robustness for scientific imaging (e.g., microscopy). We do not foresee direct negative societal impacts arising specifically from this algorithmic improvement beyond the general dual-use risks associated with standard computer vision capabilities.
    
\item {\bf Safeguards}
    \item[] Question: Does the paper describe safeguards that have been put in place for responsible release of data or models that have a high risk for misuse (e.g., pre-trained language models, image generators, or scraped datasets)?
    \item[] Answer: \answerNA{} 
    \item[] Justification: The paper does not release high-risk models (e.g., pre-trained LLMs or image generators) or internet-scraped datasets.

\item {\bf Licenses for existing assets}
    \item[] Question: Are the creators or original owners of assets (e.g., code, data, models), used in the paper, properly credited and are the license and terms of use explicitly mentioned and properly respected?
    \item[] Answer: \answerYes{} 
    \item[] Justification: We properly cite all existing datasets used (CIFAR-10, ImageNet-100, BBBC048) and the off-the-shelf denoisers (e.g., DRUNet) and baseline methods.

\item {\bf New assets}
    \item[] Question: Are new assets introduced in the paper well documented and is the documentation provided alongside the assets?
    \item[] Answer: \answerNA{} 
    \item[] Justification: The paper does not introduce or release any new datasets.

\item {\bf Crowdsourcing and research with human subjects}
    \item[] Question: For crowdsourcing experiments and research with human subjects, does the paper include the full text of instructions given to participants and screenshots, if applicable, as well as details about compensation (if any)? 
    \item[] Answer: \answerNA{} 
    \item[] Justification: The research does not involve crowdsourcing or human subjects.

\item {\bf Institutional review board (IRB) approvals or equivalent for research with human subjects}
    \item[] Question: Does the paper describe potential risks incurred by study participants, whether such risks were disclosed to the subjects, and whether Institutional Review Board (IRB) approvals (or an equivalent approval/review based on the requirements of your country or institution) were obtained?
    \item[] Answer: \answerNA{} 
    \item[] Justification: The research does not involve human subjects.

\item {\bf Declaration of LLM usage}
    \item[] Question: Does the paper describe the usage of LLMs if it is an important, original, or non-standard component of the core methods in this research? Note that if the LLM is used only for writing, editing, or formatting purposes and does \emph{not} impact the core methodology, scientific rigor, or originality of the research, declaration is not required.
    \item[] Answer: \answerNA{} 
    \item[] Justification: LLMs were not used as an important, original, or non-standard component of the core methods in this research.

\end{enumerate}
\end{document}